\documentclass[aps,prd,longbibliography,preprintnumbers,nofootinbib,
superscriptaddress,amsmath,floatfix]{revtex4}

\usepackage{amssymb}  % \gtrsim, \geqslant, etc: see amsguide.ps
\usepackage{graphicx,subfigure}
\usepackage{exscale}
\usepackage{textcomp}
\usepackage{enumerate}
\usepackage [english]{babel}
\usepackage[utf8]{inputenc}
\usepackage [autostyle, english = american]{csquotes}
\MakeOuterQuote{"}
\usepackage{color}

\usepackage[normalem]{ulem}  % \sout{old text} for strikeout

\newcommand{\non}{\nonumber\\}

\newcommand{\be}{\begin{equation}}
\newcommand{\ee}{\end{equation}}
\newcommand{\bea}{\begin{eqnarray}}
\newcommand{\eea}{\end{eqnarray}}
\newcommand{\ba}[1]{\begin{array}{#1}}
\newcommand{\ea}{\end{array}}

\newcommand{\Tr}{{\rm Tr}}

\begin{document}

%%%%%%%%%%%%%%%%%%%%%%%%%%%%%%%%%%%%%%%%%%%%%%%%%%%%%%%%%%%%%%%%%%%%%%%
\title{Strange quark mass turns magnetic domain walls into multi-winding flux tubes}

\author{Geraint W.\ Evans}
\email{g.w.evans@soton.ac.uk}
\affiliation{Mathematical Sciences and STAG Research Centre, University of Southampton, Southampton SO17 1BJ, United Kingdom}

\author{Andreas Schmitt}
\email{a.schmitt@soton.ac.uk}
\affiliation{Mathematical Sciences and STAG Research Centre, University of Southampton, Southampton SO17 1BJ, United Kingdom}
%%%%%%%%%%%%%%%%%%%%%%%%%%%%%%%%%%%%%%%%%%%%%%%%%%%%%%%%%%%%%%%%%%%%%%%

\date{4 December 2020}

%%%%%%%%%%%%%%%%%%%%%%%%%%%%%%%%%%%%%%%%%%%%%%%%%%%%%%%%%%%%%%%%%%%%%%%
\begin{abstract} 
Dense quark matter is expected to behave as a type-II superconductor at strong coupling. It was previously shown that if the strange quark mass $m_s$ is neglected, magnetic domain walls in the so-called 2SC phase are the energetically preferred magnetic defects in a certain parameter region. Computing the flux tube profiles and associated free energies within a Ginzburg-Landau approach, we find a cascade of multi-winding flux tubes as "remnants" of the domain wall when $m_s$ is increased. These flux tubes exhibit an unconventional ring-like 
structure of the magnetic field. We show that flux tubes with winding numbers larger than one survive for values of $m_s$ up to about 20\% of the quark chemical potential. This makes them unlikely to play a significant role in compact stars, but they may appear in the QCD phase diagram in the presence of an external magnetic field. 
%\vspace{1.5cm}
\end{abstract}
%%%%%%%%%%%%%%%%%%%%%%%%%%%%%%%%%%%%%%%%%%%%%%%%%%%%%%%%%%%%%%%%%%%%%%%

\maketitle

%\tableofcontents

%%%%%%%%%%%%%%%%%%%%%%%%%%%%%%%%%%%%%%%%%%%%%%%%%%%%%%%%%%%%%%%
\section{Introduction}
%%%%%%%%%%%%%%%%%%%%%%%%%%%%%%%%%%%%%%%%%%%%%%%%%%%%%%%%%%%%%%%

Cold and dense matter is a color superconductor, in which certain color-magnetic fields are screened just like ordinary magnetic fields are screened in an electronic superconductor \cite{Alford:2007xm}. What happens if a color superconductor is placed in an external {\it ordinary} magnetic field? This question is, firstly, of theoretical interest because it addresses the phase structure of Quantum Chromodynamics (QCD). Especially for vanishing baryon density, where lattice calculations can be employed, the behavior of QCD in an external magnetic field has caught a lot of attention \cite{Bali:2011qj,Kharzeev:2013jha}, motivated by the creation of large magnetic fields in heavy-ion collisions \cite{Kharzeev:2007jp,Adamczyk:2015eqo}. Secondly, the question is of phenomenological interest because color-superconducting quark matter may be found in the interior of neutron stars or in quark stars. Some of these compact stars (then called magnetars) are known to have surface magnetic fields up to about $10^{15}\, {\rm G}$. Effects on the QCD phase structure can be expected to occur for magnetic fields of the order of  $10^{18}\, {\rm G}$ and higher \cite{Kharzeev:2013jha}. This is compatible with  
estimates for the maximal possible magnetic fields in the interior of the star \cite{1991ApJ...383..745L,Ferrer:2010wz,Potekhin:2011eb}, although explicit calculations suggest that it is unlikely that the field in the core increases by several orders of magnitude from surface to core \cite{Bocquet:1995je}.  Possibly, although  perhaps less likely, color superconductors may also be created in future collider experiments that aim to reach large baryon densities \cite{Schmitt:2016pre}. 
How a color superconductor responds to an external 
ordinary magnetic field strongly depends on the particular color-superconducting phase. 

In this paper we mostly focus on the so-called 2SC phase
\cite{Bailin:1983bm}, where only up and down quarks participate in Cooper pairing, while the strange quarks and all quarks of one color remain unpaired. In this phase, just like in the color-flavor-locked (CFL) phase \cite{Alford:1998mk}, there is a certain combination of the photon and the gluons whose magnetic field penetrates the color superconductor unperturbed \cite{Schmitt:2003aa}. In other words, the Cooper pairs are neutral with respect to a "rotated charge", which is a combination of electromagnetic and color charges. As a consequence, a certain fraction of an external ordinary magnetic field can penetrate in the form of this rotated field. It turns out that at strong coupling (strong coupling constant much greater than the electromagnetic one) this fraction is in fact very large, such that only a small fraction of the magnetic field is expelled. Nevertheless, it may be energetically favorable to admit additional ordinary magnetic flux through certain magnetic defects. Another way of saying this is that there is a combination of the photon and the gluons whose magnetic component is completely expelled for small external magnetic field but (partially) penetrates the color superconductor for sufficiently large external fields. As in electronic superconductors, this scenario is referred to as type-II superconductivity.  We expect the 2SC and CFL phases to be in the type-II regime for sufficiently large pairing gaps, which are estimated to be reached for realistic coupling strengths under compact star conditions \cite{Iida:2002ev,Iida:2004if,Alford:2010qf,Haber:2017oqb}.  

The main idea of this paper is as follows. If the strange quark mass $m_s$ is neglected, then the usual 2SC phase is indistinguishable from the phase where only up and strange quarks form Cooper pairs. To avoid 
confusion, we refer to these  phases as 2SC$_{\rm ud}$ and 2SC$_{\rm us}$, respectively (the third possibility, 2SC$_{\rm ds}$, is different even for $m_s\simeq 0$ because of the electric charges of the quarks). This enhanced flavor symmetry renders a planar domain wall configuration  possible, 
which smoothly interpolates from 2SC$_{\rm ud}$ to 2SC$_{\rm us}$. Indeed, there is a parameter regime where, due to the admission of additional magnetic flux through the wall, the formation of such domain walls becomes energetically favorable \cite{Haber:2017oqb}\footnote{These domain walls are different from the 2SC domain walls associated with the spontaneous breaking of the axial $U(1)_A$ \cite{Son:2007ny}, which are perpendicular, not parallel, to the magnetic field and which have nothing to do with the rotation from the $ud$ to the $us$ condensate.}. At sufficiently large densities, $m_s$ can safely be neglected compared to the chemical potential, but this is no longer true in the context of compact stars (where we may still neglect the masses of up and down quarks). In this case,  the free energies of 2SC$_{\rm ud}$ and 2SC$_{\rm us}$ phases become non-degenerate because there is an energy cost involved in including the massive strange quarks in the Cooper pair condensate. Therefore, the two phases can no longer coexist on two sides of a planar defect. 
As we shall see, the planar defect turns into a flux tube whose radius decreases with increasing $m_s$. This is realized in a descending sequence of multi-winding flux tubes, until we are left with flux tubes with winding number one at sufficiently large values of $m_s$. The flux tubes in this sequence are 2SC$_{\rm ud}$ flux tubes with a 2SC$_{\rm us}$ core. Flux tubes with a core different from normal-conducting matter are possible in multi-component systems such as quark matter, for instance CFL flux tubes with a 2SC core, which are preferred over CFL flux tubes with a normal core for realistic parameters 
\cite{Haber:2017oqb}. Related examples are two-component mixtures of superconductors with superfluids, where the core of the flux tubes is normal-conducting, but the superfluid condensate remains nonzero \cite{Alford:2007np,Haber:2016ljn,Haber:2017kth}. Our flux tubes are different from these examples in that the induced condensate in the core is zero far away from the center of the flux tube. They also differ from these cases with regard to the multi-winding solutions: While it has been pointed out that multi-winding flux tubes can become favored in CFL matter \cite{Haber:2018tqw}
and in superconductor/superfluid mixtures \cite{Alford:2007np}, this concerns a small region close to the transition point between type-I and type-II behavior of the superconductor, whose precise phase structure is further complicated by a first-order transition due to an attractive long-range interaction between the flux tubes \cite{Iida:2004if,Buckley:2004ca,Haber:2017kth,Haber:2017oqb}. The multi-winding 2SC flux tubes discussed here exist in a much larger parameter region within the type-II regime.

We shall work within a Ginzburg-Landau framework, which has often been used to study certain aspects of color superconductivity \cite{Blaschke:1999fy,Iida:2000ha,Iida:2001pg,Sedrakian:2002mk,Iida:2002ev,Giannakis:2003am,Hatsuda:2006ps}. In particular, our starting point is a direct extension of Ref.\ \cite{Haber:2017oqb}, making use of mass insertions considered previously  \cite{Iida:2003cc,Iida:2004cj,Schmitt:2010pf}. While the Ginzburg-Landau approach in principle allows us to make model-independent predictions, it also has several shortcomings. Perhaps most importantly, it is purely bosonic and thus does not account for the fermionic constituents of the Cooper pairs. This is relevant since the magnetic fields considered here are large enough to potentially modify the microscopic structure of the Cooper pair condensates, as demonstrated in fermionic frameworks \cite{Ferrer:2005vd,Ferrer:2006vw,Noronha:2007wg,Fukushima:2007fc,Yu:2012jn}. Moreover, for our numerical evaluation we shall employ the weak-coupling form of the Ginzburg-Landau parameters and extrapolate the results to large  values of the strong coupling constant. Also, we do not attempt to construct any flux tube arrays, but rather focus on single flux tubes. This enables us to compute the critical magnetic field at which it becomes favorable to place a single magnetic flux tube into the system, $H_{c1}$ in the usual terminology, but we cannot determine the precise structure of the phase above this critical magnetic field. Finally, we only include the effect of the quark mass to lowest nontrivial order and thus for large quark masses our results have to be considered with some care. 

We emphasize that the flux tubes considered here are "pure" magnetic flux tubes, i.e., they have (quantized) magnetic flux, but zero baryon circulation. In color-superconducting matter, such flux tubes are not protected by topology \cite{Alford:2010qf,Eto:2013hoa}. We shall compute the parameter regime where they become {\it energetically} stable, i.e., where they cannot decay despite their non-topological nature.  Since the 2SC phase is not a superfluid, pure magnetic flux tubes are the only possible line defects. For comparison, the CFL phase allows for defects with  nonzero magnetic flux {\it and} nonzero baryon circulation  \cite{Eto:2009kg,Alford:2016dco}. As for ordinary superfluid vortices, their energy is formally divergent, while the magnetic flux tubes considered here have finite energy, just like magnetic flux tubes in an ordinary electronic superconductor.

The main difference of our magnetic flux tubes compared to the textbook scenario is the appearance of multiple condensates and gauge fields. Therefore, our study can also be put into the wider context of magnetic defects in unconventional superconductors.  By considering the diagonal subsector of the order parameter in color-flavor space, we allow for three nonzero condensates, and as a consequence we can restrict  ourselves to three gauge fields of the color and electromagnetic gauge group $SU(3)\times U(1)$. This is somewhat similar to electroweak strings with gauge group $SU(2)\times U(1)$ \cite{Vachaspati:1991dz,Achucarro:1999it,Volkov:2006ug}. In the so-called semilocal approximation, where the $SU(2)$ remains ungauged, these strings are described within an abelian Higgs model  \cite{Chernodub:2010sg}. In this context, the flux tube profiles are often calculated at the transition point between type-I and type-II superconductivity, also referred to as the Bogomolny limit \cite{bogomolny}. Our calculation is not restricted to this point and in fact we shall see that the multi-winding flux tubes are only stable away from this transition point.  As in Ref.\ \cite{Chernodub:2010sg} we will see that for the multi-winding flux tubes the profiles of the magnetic field are ring-like, i.e., the maximum of the magnetic field is not at the center of the flux tube. This ring-like structure has also been observed in a model with a non-standard magnetic permeability \cite{Bazeia:2018hyv}, and it is similar to the  experimentally observed structure of two-component superfluid vortices  \cite{PhysRevLett.83.2498}.

Our paper is organized as follows. We start with setting up the Ginzburg-Landau formalism in Sec.\ \ref{sec:setup}, including the discussion of the mass terms and the rotated electromagnetism. In Sec.\ \ref{sec:hom} we discuss the homogeneous phases as a preparation for the subsequent sections. The critical magnetic field $H_c$ and the upper critical field $H_{c2}$ are computed in Sec.\ \ref{sec:HHH}, and we set up the calculation of the flux tube profiles in Sec.\ \ref{sec:Hc1} needed for the numerical calculation of the lower critical field $H_{c1}$. The main results are presented and discussed in Sec.\ \ref{sec:results}, which is divided into the discussion of the flux tubes themselves, Sec.\ \ref{sec:tubes}, and the resulting phase diagram, Sec.\ \ref{sec:phase}. We give a summary and an outlook in Sec.\ \ref{sec:summary}. 
Our convention for the metric tensor is $g^{\mu\nu}={\rm diag}(1,-1,-1,-1)$. We work in natural units $\hbar=c=k_B=1$ and use Heaviside-Lorentz units for the gauge fields, in which the elementary charge is $e=\sqrt{4\pi\alpha}\simeq 0.3$, where $\alpha$ is the fine-structure constant.

%%%%%%%%%%%%%%%%%%%%%%%%%%%%%%%%%%%%%%%%%%%%%%%%%%%%%%%%%%%%%%%
\section{Setup}
\label{sec:setup}
%%%%%%%%%%%%%%%%%%%%%%%%%%%%%%%%%%%%%%%%%%%%%%%%%%%%%%%%%%%%%%%

%%%%%%%%%%%%%%%%%%%%%%%%%%%%%%%%%%%%%%%%%%%%%%%%%%%%%%%%%%%%%%%
\subsection{Ginzburg-Landau potential}
\label{sec:GL}
%%%%%%%%%%%%%%%%%%%%%%%%%%%%%%%%%%%%%%%%%%%%%%%%%%%%%%%%%%%%%%%

In three-flavor quark matter with sufficiently small mismatch in the Fermi momenta of the different fermion species, Cooper pairing predominantly occurs in the spin-zero channel and in the antisymmetric anti-triplet channels in color and flavor space, $[\bar{3}]_c$ and $[\bar{3}]_f$. As pairing is assumed to occur between fermions of the same chirality, the flavor channel stands for either left-handed or right-handed fermions. Therefore, the order parameter for Cooper pair condensation can be written as 
\be \label{Psi}
\Psi = \Phi_{ij} J_i\otimes I_j \in [\bar{3}]_c \otimes [\bar{3}]_f \, , 
\ee
where the anti-symmetric $3\times 3$ matrices  $(J_i)_{jk} = -i\epsilon_{ijk}$ and $(I_j)_{k\ell} = -i\epsilon_{jk\ell}$ form bases of the three-dimensional spaces $[\bar{3}]_c$ and $[\bar{3}]_f$, respectively. As a consequence, we can characterize a color-superconducting phase by the $3\times 3$ matrix $\Phi$, which has one (anti-)color and one (anti-)flavor index. We shall put the colors in the order $(r, g, b)$ (for red, green, blue) and the flavors in the order $(u, d, s)$ (for up, down, strange). The color charges are only labels and thus their order is not crucial, but the order of the flavors matters since electric charge and quark masses break the flavor symmetry\footnote{For the CFL phase the order $(d,s,u)$ is often used, which is convenient since then the generator of the electromagnetic gauge group is proportional to the eighth generator of the color gauge group. Since our main focus is on the 2SC phase, this re-ordering is not necessary and we work with the more common order $(u,d,s)$.}. In this convention, for instance, $\Phi_{11}$ carries anti-indices $\bar{r}$ and $\bar{u}$ and thus describes pairing of $gd$ with $bs$ quarks and of $gs$ with $bd$ quarks. 

We consider the following Ginzburg-Landau potential up to fourth order in $\Phi$,
\bea \label{UPhi}
U &=& -12\Big\{\Tr[(D_0\Phi)^\dag(D_0\Phi)]+u^2\Tr[(D_i\Phi)^\dag(D^i\Phi)]\Big\}-4\mu^2 \Tr[\Phi^\dag\Phi]+4\sigma\,\Tr\left[\Phi^\dag \left(\frac{M^2}{2\mu_q}+\mu_eQ\right)\Phi\right]\non[2ex]
&&+16(\lambda+h)\Tr[(\Phi^\dag\Phi)^2]-16h (\Tr[\Phi^\dag\Phi])^2 +\frac{1}{4}F_{\mu\nu}^aF_a^{\mu\nu}+\frac{1}{4}F_{\mu\nu}F^{\mu\nu}
 \, . 
\eea
Apart from the mass correction, proportional to the parameter $\sigma$, this is exactly the same potential, and the same notation, as in Ref.\ \cite{Haber:2017oqb}, where the starting point was the potential for $\Psi$, based on previous works \cite{Iida:2000ha,Iida:2001pg,Iida:2002ev,Giannakis:2003am}. Due to the broken Lorentz invariance in the medium, the temporal and spatial parts of the kinetic term have different prefactors, $u=1/\sqrt{3}$. The covariant derivative is given by 
\be
D_\mu \Phi = \partial_\mu\Phi-ig A_\mu^a T_a^T\Phi+ieA_\mu\Phi Q \, ,
\ee 
where $g$ and $e$ are the strong and electromagnetic coupling constants, respectively. The color gauge fields are denoted by $A_\mu^a$, $a=1,\ldots, 8$, and $A_\mu$ is the electromagnetic gauge field. Furthermore, $T_a=\lambda_a/2$, with the Gell-Mann matrices $\lambda_a$, are the generators of the color gauge group $SU(3)$, and the electric charge matrix for the Cooper pairs $Q = {\rm diag}(q_d+q_s,q_u+q_s,q_u+q_d) = {\rm diag}(-2/3,1/3,1/3)$ is the generator of the electromagnetic gauge group $U(1)$. Here, $q_u$, $q_d$, $q_s$ denote the individual quark charges in units of the elementary charge $e$. The field strength tensors are  $F_{\mu\nu}^a = \partial_\mu A_\nu^a-\partial_\nu A_\mu^a+gf^{abc}A_\mu^bA_\nu^c$ for the color sector, where $f^{abc}$ are the $SU(3)$ structure constants,  and $F_{\mu\nu} = \partial_\mu A_\nu-\partial_\nu A_\mu$ for the electromagnetic sector. The constants in front of the quadratic and quartic terms in $\Phi$ are written conveniently as combinations of $\mu$, $\lambda$ and $h$, whose physical meaning will become obvious after performing the traces (see Eq.\ (\ref{U1})).

We have incorporated a mass correction to lowest order, with the (fermionic) quark chemical potential $\mu_q$ and the mass matrix for the Cooper pairs $M={\rm diag}(m_d+m_s,m_u+m_s,m_u+m_d)\simeq {\rm diag}(m_s,m_s,0)$, i.e., we shall neglect the masses of the light quarks, $m_u\simeq m_d\simeq 0$, and keep the strange quark mass $m_s$ as a free parameter. We have also included the contribution of the electric charge chemical potential $\mu_e$, which is of the same order for small quark masses, $\mu_e\propto m_s^2/\mu_q$. The correction term is identical to the one used in Refs.\ \cite{Iida:2003cc,Iida:2004cj}, which is easily seen by an appropriate rescaling of $\Phi$. 

In the following we restrict ourselves to diagonal order parameters, $\Phi=\frac{1}{2}{\rm diag}(\phi_1,\phi_2,\phi_3)$
with $\phi_i\in \mathbb{C}$, where $\phi_1$ corresponds to $ds$ pairing, $\phi_2$ to $us$ pairing, and $\phi_3$ to $ud$ pairing. 
If flavor symmetry was intact, off-diagonal order parameters could always be brought into a diagonal form by an appropriate rotation in color and flavor space. This is no longer true when flavor symmetry is explicitly broken, and thus our restriction to diagonal order parameters is a simplification of the most general situation \cite{Rajagopal:2005dg}. Even in this simplified case, $2^3=8$ qualitatively different homogeneous phases have to be considered
in principle, accounting for each of the three condensates to be either zero or nonzero. The restriction to diagonal matrices $\Phi$ allows us to consistently set all gauge fields with off-diagonal components to zero, $A_\mu^1=A_\mu^2=A_\mu^4=A_\mu^5=A_\mu^6=A_\mu^7=0$, such that the only relevant gauge fields are the two color gauge fields $A_\mu^3$, $A_\mu^8$, and the electromagnetic gauge field $A_\mu$. Moreover, we are only interested in static solutions and drop all electric fields, i.e., we only keep the spatial components of the gauge fields, giving rise to the magnetic fields ${\bf B}_3 = \nabla\times {\bf A}_3$,  ${\bf B}_8 = \nabla\times {\bf A}_8$, and ${\bf B} = \nabla\times {\bf A}$.

Within this ansatz, performing the traces in Eq.\ (\ref{UPhi}) yields
\bea \label{U1}
U &=& \left|\left(\nabla+\frac{ig{\bf A}_3}{2}+\frac{ig{\bf A}_8}{2\sqrt{3}}+\frac{2ie{\bf A}}{3}\right)\phi_1\right|^2 +\left|\left(\nabla-\frac{ig{\bf A}_3}{2}+\frac{ig{\bf A}_8}{2\sqrt{3}}-\frac{ie{\bf A}}{3}\right)\phi_2\right|^2  + \left|\left(\nabla-\frac{ig{\bf A}_8}{\sqrt{3}}-\frac{ie{\bf A}}{3}\right)\phi_3\right|^2 \non[2ex]
&&-(\mu^2-m_1^2)|\phi_1|^2-(\mu^2-m_2^2)|\phi_2|^2-(\mu^2-m_3^2)|\phi_3|^2 +\lambda(|\phi_1|^4+|\phi_2|^4+|\phi_3|^4) \non[2ex]
&&-2h(|\phi_1|^2|\phi_2|^2+|\phi_1|^2|\phi_3|^2+|\phi_2|^2|\phi_3|^2)+\frac{{\bf B}_3^2}{2}+\frac{{\bf B}_8^2}{2}+\frac{{\bf B}^2}{2} \, .
\eea
This potential can be viewed as a generalized version of a textbook superconductor which has a single condensate coupled to a single gauge field, see for instance Ref.\ \cite{tinkham2004introduction}. Here we have three condensates with identical (bosonic) chemical potential $\mu$, with self-coupling $\lambda$, cross-coupling $h$, and three different effective masses (squared)
\bea \label{m123}
m_1^2 &=&  \sigma\left(\frac{m_s^2}{2\mu_q}-\frac{2\mu_e}{3}\right)     \, , \qquad 
m_2^2 =  \sigma\left(\frac{m_s^2}{2\mu_q}+\frac{\mu_e}{3}\right)  \, , \qquad 
m_3^2 = \sigma\frac{\mu_e}{3} \, .
\eea
All three gauge fields couple to the condensates. We can simplify the potential by a suitable rotation of the gauge fields.

%%%%%%%%%%%%%%%%%%%%%%%%%%%%%%%%%%%%%%%%%%%%%%%%%%%%%%%%%%%%%%%
\subsection{Rotated electromagnetism and Gibbs free energy}
\label{sec:rotate}
%%%%%%%%%%%%%%%%%%%%%%%%%%%%%%%%%%%%%%%%%%%%%%%%%%%%%%%%%%%%%%%

We apply the double rotation from Ref.\ \cite{Haber:2017oqb}, which, denoting the rotated gauge fields by $\tilde{A}_\mu^3$, $\tilde{A}_\mu^8$, and $\tilde{A}_\mu$, reads, 
\be \label{rotatetwice}
\left(\begin{array}{c} \tilde{A}_\mu^3 \\ \tilde{A}_\mu^8 \\ \tilde{A}_\mu \end{array}\right) =  
\left(\begin{array}{ccc} \cos\vartheta_2 & 0 &\sin\vartheta_2 \\ 0&1&0 \\ -\sin\vartheta_2&0&\cos\vartheta_2  \end{array}\right)\left(\begin{array}{ccc} 1 & 0 &0 \\ 0& \cos\vartheta_1 & \sin\vartheta_1 \\ 0 & -\sin\vartheta_1 & \cos\vartheta_1   \end{array}\right)\left(\begin{array}{c} A_\mu^3 \\ A_\mu^8 \\ A_\mu \end{array}\right)  \, ,
\ee
where the mixing angles $\vartheta_1$ and $\vartheta_2$ are given by
\begin{subequations} \label{thetamix}
\bea
\sin\vartheta_1 &=& \frac{e}{\sqrt{3g^2+e^2}} \, , \qquad \cos\vartheta_1 = \frac{\sqrt{3}g}{\sqrt{3g^2+e^2}} \, , \\[2ex]
\sin\vartheta_2 &=& \frac{\sqrt{3}e}{\sqrt{3g^2+4e^2}} \, , \qquad \cos\vartheta_2 = \frac{\sqrt{3g^2+e^2}}{\sqrt{3g^2+4e^2}} \, .
\eea
\end{subequations}
This rotation is the most convenient choice for our purpose of calculating flux tube profiles in the 2SC phase: The first rotation, with the usual "2SC mixing angle" $\vartheta_1$, ensures that in the homogeneous 2SC phase, where only the condensate $\phi_3$ is nonzero, only the $\tilde{\bf B}_8$ field is expelled. The other two rotated fields penetrate the superconductor unperturbed (assuming zero magnetization from the unpaired quarks). If we were only interested in the homogeneous 2SC phase, this rotation would be sufficient. However, we will allow for $\phi_1$ and $\phi_2$ to be induced in the core of the flux tube. Therefore, we apply a second rotation with mixing angle $\vartheta_2$. This rotation leaves $\tilde{\bf B}_8$ invariant and creates a field, namely $\tilde{\bf B}$, which is unaffected by the superconductor even if all three condensates are nonzero. Thus, $\tilde{\bf B}$ simply decouples from the condensates and can be ignored in the  calculation of the flux tube profiles. For $g\gg e$ both mixing angles are small and thus in this case $\tilde{A}_\mu^3$ and $\tilde{A}_\mu^8$ are "almost" gluons with a small admixture of the photon, while $\tilde{A}_\mu$ is "almost" the photon with a small gluonic admixture. For consistency, we shall work with the rotated fields (\ref{rotatetwice}) throughout the paper, including the discussion of the homogeneous phases in Sec.\ \ref{sec:hom}.  

Applying the gauge field rotation and writing the complex fields in terms of their moduli and phases,
\be
\phi_i({\bf r}) = \frac{\rho_i({\bf r})}{\sqrt{2}}e^{i\psi_i({\bf r})} \, , \qquad i =1,2,3,
\ee
the Ginzburg-Landau potential (\ref{U1}) becomes
\be \label{UU0}
U = U_0 +\frac{\tilde{\bf B}_3^2}{2}+\frac{\tilde{\bf B}_8^2}{2} +\frac{\tilde{\bf B}^2}{2} \, , 
\ee
with 
\bea \label{U0}
U_0 
&=&\left(\nabla\psi_1+\tilde{q}_3\tilde{{\bf A}}_3+\tilde{q}_{81}\tilde{{\bf A}}_8\right)^2\frac{\rho_1^2}{2}
+\left(\nabla\psi_2-\tilde{q}_3\tilde{{\bf A}}_3+\tilde{q}_{82}\tilde{{\bf A}}_8\right)^2\frac{\rho_2^2}{2}
+\left(\nabla\psi_3-\tilde{q}_{83}\tilde{{\bf A}}_8\right)^2\frac{\rho_3^2}{2} \non[2ex]
&&+\frac{(\nabla\rho_1)^2}{2}+\frac{(\nabla\rho_2)^2}{2}+\frac{(\nabla\rho_3)^2}{2}
-\frac{\mu^2-m_1^2}{2}\rho_1^2-\frac{\mu^2-m_2^2}{2}\rho_2^2-\frac{\mu^2-m_3^2}{2}\rho_3^2 +\frac{\lambda}{4}(\rho_1^4+\rho_2^4+\rho_3^4)\non[2ex]
&&-\frac{h}{2}(\rho_1^2\rho_2^2+\rho_1^2\rho_3^2+\rho_2^2\rho_3^2)  \, , 
\eea
where we have introduced the rotated charges
\begin{subequations}
\bea
\tilde{q}_{81}&\equiv& \frac{g}{2\sqrt{3}}\cos\vartheta_1+\frac{2e}{3}\sin\vartheta_1 = \frac{3g^2+4e^2}{6\sqrt{3g^2+e^2}} \, , \\[2ex]
\tilde{q}_{82}&\equiv& \frac{g}{2\sqrt{3}}\cos\vartheta_1-\frac{e}{3}\sin\vartheta_1 = \frac{3g^2-2e^2}{6\sqrt{3g^2+e^2}} \, , \\[2ex]
\tilde{q}_{83}&\equiv& \frac{g}{\sqrt{3}}\cos\vartheta_1+\frac{e}{3}\sin\vartheta_1 = \frac{\sqrt{3g^2+e^2}}{3} \, , \\[2ex]
\tilde{q}_3 &\equiv& \frac{g}{2}\cos\vartheta_2+\frac{e}{2}\cos\vartheta_1\sin\vartheta_2= \frac{g}{2}\frac{\sqrt{3g^2+4e^2}}{\sqrt{3g^2+e^2}} \, .
\eea
\end{subequations}
In Eq.\ (\ref{UU0}) we have separated the quadratic contributions of the magnetic fields, which is notationally convenient for the following. 

We shall be interested in the phase structure at fixed external magnetic field ${\bf H}$, which we assume to be homogeneous and along the $z$-direction, ${\bf H}({\bf r})=H{\bf e}_z$. Therefore, we need to consider the Gibbs free energy density
\bea
G &=& \frac{1}{V} \int d^3{\bf r}\,(U -{\bf H}\cdot{\bf B}) \, ,
\eea
where $V$ is the volume of our system. We can obviously assume that all induced magnetic fields have only $z$-components as well. Denoting the $z$-components of the rotated fields by $\tilde{B}_3$, $\tilde{B}_8$, and $\tilde{B}$, we have  ${\bf H}\cdot{\bf B} = H[\sin\vartheta_1\, \tilde{B}_8+\cos\vartheta_1(\sin\vartheta_2\, \tilde{B}_3+\cos\vartheta_2\,\tilde{B})]$. Since $\tilde{\bf B}$ does not couple to any of the condensates, it remains homogeneous even in the presence of flux tubes. Consequently, the equation of motion for $\tilde{\bf A}$ is trivially fulfilled, and we determine $\tilde{B}$ by minimizing the Gibbs free energy, which, using Eq.\ (\ref{UU0}), yields
\be \label{tildeB} 
\tilde{B} =H\cos\vartheta_1\cos\vartheta_2 \, .
\ee 
Reinserting this result into $G$, we obtain 
\bea\label{gibbs}
G&=& -\frac{H^2\cos^2\vartheta_1\cos^2\vartheta_2}{2}+\frac{1}{V}\int d^3{\bf r}\,\left[U_0 +\frac{\tilde{B}_3^2}{2}+\frac{\tilde{B}_8^2}{2}-H(\sin\vartheta_1\, \tilde{B}_8+\cos\vartheta_1\sin\vartheta_2\, \tilde{B}_3)\right] \, .
\eea

%%%%%%%%%%%%%%%%%%%%%%%%%%%%%%%%%%%%%%%%%%%%%%%%%%%%%%%%%%%%%%%
\subsection{Parameter choices}
\label{sec:para}
%%%%%%%%%%%%%%%%%%%%%%%%%%%%%%%%%%%%%%%%%%%%%%%%%%%%%%%%%%%%%%%

The potential $U_0$ (\ref{U0}) depends on the parameters $\mu$, $\lambda$, $h$, $\sigma$. The discussion of the homogeneous phases in Sec.\ \ref{sec:hom} turns out to be sufficiently simple to keep these parameters unspecified and to investigate the general phase structure. Our main results, however, require the numerical calculation of the flux tube profiles, and a completely general study would be extremely laborious. Therefore, for the results in Sec.\ \ref{sec:results}, we employ the weak-coupling values of these parameters \cite{Iida:2002ev,Giannakis:2003am,Iida:2003cc,Iida:2004cj,Iida:2004if,Eto:2013hoa}, 
\bea \label{weak}
\mu^2 &=& \frac{48\pi^2}{7\zeta(3)}T_c^2\left(1-\frac{T}{T_c}\right) \, , \qquad 
\lambda = \frac{72\pi^4}{7\zeta(3)}\frac{T_c^2}{\mu_q^2} \, , \qquad 
h= -\frac{36\pi^4}{7\zeta(3)}\frac{T_c^2}{\mu_q^2} \, , \qquad 
\sigma = -\frac{24\pi^2}{7\zeta(3)}\frac{T_c^2}{\mu_q}\ln\frac{T_c}{\mu_q} \, , 
\eea
where $T_c$ is the critical temperature. The ratio $T_c/\mu_q$ can be understood as a measure of the pairing strength since $T_c$ is closely related to the pairing gap. For instance, at weak coupling, which is applicable at asymptotically large densities, the zero-temperature pairing gap is exponentially suppressed compared to $\mu_q$. It is related by a numerical factor of order one to $T_c$ \cite{Schmitt:2002sc}, and thus $T_c/\mu_q$ is also exponentially small.  We shall extrapolate our results to strong coupling, having in mind applications to compact stars, where the densities are large, but not asymptotically large. In this case, model calculations as well as extrapolations of perturbative results suggest that $T_c$ is of the order of tens of MeV, while we expect $\mu_q \sim (400 - 500)\, {\rm MeV}$ in the cores of compact stars, which yields
the estimate $T_c/\mu_q \sim 0.1$. Besides the 
implicit dependence on $T_c/\mu_q$ our potential also depends on the ratio $m_s/\mu_q$. Since $m_s$ is medium dependent, its value at non-asymptotic densities is poorly known. It is expected to be somewhere between the current mass and the constituent mass within a baryon, $m_s \sim (100 - 500)\, {\rm MeV}$; for a concrete calculation within the Nambu--Jona-Lasinio model see for instance Ref.\ \cite{Ruester:2005jc}. With the range of the quark chemical potential given above  we thus expect $m_s/\mu_q \sim (0.2 - 1)$. Finally, our potential depends on the electric charge chemical potential $\mu_e$. In a fermionic approach, this chemical potential would be determined from the conditions of beta-equilibrium and charge neutrality. Since our Ginzburg-Landau expansion is formally based on small values of the order parameter, we follow Refs.\ \cite{Iida:2003cc,Iida:2004cj} and use the value of $\mu_e$ in the completely unpaired phase. At weak coupling and to lowest order in the strange quark mass this value is (see for instance Refs.\ \cite{Alford:2002kj,Schmitt:2010pn})
\be \label{mue}
\mu_e = \frac{m_s^2}{4\mu_q} \, .
\ee
With this result, it is convenient to trade the dimensionful parameter $\sigma$ for the dimensionless "mass parameter"
\be\label{alpha}
\alpha \equiv \frac{\sigma m_s^2}{\mu^2\mu_q} = \frac{m_s^2}{2\mu_q^2}\left(1-\frac{T}{T_c}\right)^{-1}\ln\frac{\mu_q}{T_c} \, ,
\ee
such that the complete dependence of our potential on the strange quark mass is absorbed in $\alpha$,
\be \label{malpha}
m_1^2 = \frac{\mu^2}{3}\alpha \, , \qquad m_2^2 = \frac{7\mu^2}{12}\alpha \, , \qquad m_3^2 = \frac{\mu^2}{12}\alpha \, .
\ee
We shall see that if we are only interested in homogeneous phases, the phase structure is most conveniently calculated in the space spanned by $\alpha$, $g$, the normalized dimensionless magnetic field $H/(\mu^2/\lambda^{1/2})$, and the ratio
\be \label{eta}
\eta \equiv \frac{h}{\lambda} \, .
\ee
To ensure the boundedness of our potential, we must require $\eta<1/2$. At weak coupling $\eta=-1/2$, as one can see from Eq.\ (\ref{weak}).
  Later, in our explicit calculation of the flux tube profiles and the resulting critical magnetic fields, we consider a fixed $g$ and the parameter space spanned by $H/(\mu^2/\lambda^{1/2})$, $T_c/\mu_q$, and $m_s/\mu_q$. To choose a value of $g$,  realistic for 
compact star conditions, we observe that according to the two-loop QCD beta function (which should not be taken too seriously at such low densities), $\mu_q\simeq 400\, {\rm MeV}$ corresponds to $\alpha_s\simeq 1$ and thus $g=\sqrt{4\pi\alpha_s} \simeq 3.5$. Of course, choosing such a large value for $g$ in our main results is a bold   extrapolation, given that we work with the weak-coupling parameters (\ref{weak}). Furthermore,  we shall set $T=0$ in Eq.\ (\ref{alpha}). Strictly speaking this is inconsistent
because the Ginzburg-Landau potential is an expansion in the condensates, and we use a value for $\mu_e$ (\ref{mue}) that is only valid very close to a second-order transition to the unpaired phase. Choosing a different, nonzero temperature, would not change our result qualitatively because it only enters the relation between $m_s/\mu_q$ and $\alpha$.  The definition of $\alpha$ (\ref{alpha}) shows that the mass effect is smallest for zero temperature (i.e., in this case $\alpha$ is smallest for a given $m_s/\mu_q$). Therefore, by our choice $T=0$ in Eq.\ (\ref{alpha})  we will obtain an upper limit in $m_s/\mu_q$ for the presence of multi-winding 2SC flux tubes. Any $T>0$ in Eq.\ (\ref{alpha}) will give a smaller $m_s/\mu_q$ up to which these exotic configurations exist. In any case, the temperature dependence in the present approach is somewhat simplistic to begin with because, firstly,  in a multi-component superconductor there can be different critical temperatures for the different condensates,
resulting in temperature factors different from the standard Ginzburg-Landau formalism \cite{Haber:2017kth}. And, secondly, away from the asymptotic  weak-coupling regime the phase transition becomes first order due to  gauge field fluctuations \cite{Giannakis:2004xt}, and thus at strong coupling the behavior just below the phase transition would have to be modified in a more sophisticated approach.

%%%%%%%%%%%%%%%%%%%%%%%%%%%%%%%%%%%%%%%%%%%%%%%%%%%%%%%%%%%%%%%
\section{Homogeneous phases}
\label{sec:hom}
%%%%%%%%%%%%%%%%%%%%%%%%%%%%%%%%%%%%%%%%%%%%%%%%%%%%%%%%%%%%%%%

Here we construct all possible homogeneous candidate phases within our ansatz and compute their Gibbs free energy density. 
Since we have aligned the $z$-axis with the magnetic fields, we may write the gauge fields as $\tilde{\bf A}_3=x\tilde{B}_3{\bf e}_y$, $\tilde{\bf A}_8=x\tilde{B}_8{\bf e}_y$ with constant $\tilde{B}_3$ and $\tilde{B}_8$.  Furthermore, in this section we assume the condensates $\rho_i$ and their phases $\psi_i$ to be independent of ${\bf r}$ and thus all gradient terms in Eq.\ (\ref{U0}) vanish. Then, the equations of motion for the gauge fields and the condensates become
\begin{subequations} \label{eoma}
\bea
0&=&\rho_1^2(\tilde{q}_3\tilde{B}_3+\tilde{q}_{81}\tilde{B}_8)+\rho_2^2(\tilde{q}_3\tilde{B}_3-\tilde{q}_{82}\tilde{B}_8) \, , \label{eoma3}\\[2ex]
0&=&\rho_1^2\tilde{q}_{81}(\tilde{q}_3\tilde{B}_3+\tilde{q}_{81}\tilde{B}_8)-\rho_2^2\tilde{q}_{82}(\tilde{q}_3\tilde{B}_3-\tilde{q}_{82}\tilde{B}_8)+\rho_3^2\tilde{q}_{83}^2\tilde{B}_8  \, , \label{eoma8}
\eea
\end{subequations}
and
\begin{subequations}\label{eomrho}
\bea
0&=&\rho_1\left[x^2(\tilde{q}_3\tilde{B}_3+\tilde{q}_{81}\tilde{B}_8)^2-(\mu^2-m_1^2)+\lambda\rho_1^2-h(\rho_2^2+\rho_3^2)\right]\, ,\label{eomrho1}\\[2ex]
0&=&\rho_2\left[x^2(-\tilde{q}_3\tilde{B}_3+\tilde{q}_{82}\tilde{B}_8)^2-(\mu^2-m_2^2)+\lambda\rho_2^2-h(\rho_1^2+\rho_3^2)\right]\, ,\label{eomrho2}\\[2ex]
0&=&\rho_3\left[x^2\tilde{q}_{83}^2\tilde{B}_8^2-(\mu^2-m_3^2)+\lambda\rho_3^2-h(\rho_1^2+\rho_2^2)\right]\, . \label{eomrho3}
\eea
\end{subequations}
Since here we are only interested in homogeneous phases, the terms proportional to $x^2$ have to vanish separately in each equation (unless the equation is satisfied trivially by a vanishing condensate).

The simplest case is the normal phase, where all three condensates vanish, $\rho_1=\rho_2=\rho_3=0$. Here we expect to have no induced color-magnetic fields and the external magnetic field should penetrate unperturbed, i.e., in the unrotated basis ${\bf B}_3={\bf B}_8=0$ and ${\bf B}={\bf H}$. As a check, we can derive this result within our rotated basis. First we observe that the equations of motion (\ref{eoma}) and (\ref{eomrho}) are trivially fulfilled for vanishing condensates. Then, from Eq.\ (\ref{U0}) we obtain  $U_0=0$. Inserting this into the Gibbs free energy density (\ref{gibbs}) and minimizing the result with respect to
$\tilde{B}_3$ and $\tilde{B}_8$ yields nonzero results for these rotated magnetic fields. By undoing the rotation one can check that 
these results together with the result for $\tilde{B}$ (\ref{tildeB}) indeed give ${\bf B}_3={\bf B}_8=0$ and ${\bf B}={\bf H}$, as expected. Then, substituting the magnetic fields at the minimum back into the Gibbs free energy density yields
\be \label{GNOR}
G_{\rm NOR} = -\frac{H^2}{2} \, .
\ee
This result does not receive any mass corrections since, within our Ginzburg-Landau approach, unpaired fermions and any possible mass effects on them do not appear explicitly; the normal phase is the "vacuum" of our theory. We will now discuss the various superconducting candidate phases, which all do receive mass corrections. These mass corrections can be written in a general form using the masses $m_1$, $m_2$, $m_3$. However, especially for the free energies this leads to some lengthy expressions, which are not particularly instructive. Thus we make use of Eq.\ (\ref{malpha}) and express all free energies in this section to linear order in the mass parameter $\alpha$. 

%%%%%%%%%%%%%%%%%%%%%%%%%%%%%%%%%%%%%%%%%%%%%%%%%%%%%%%%%%%%%%%
\subsection{CFL phase} 
%%%%%%%%%%%%%%%%%%%%%%%%%%%%%%%%%%%%%%%%%%%%%%%%%%%%%%%%%%%%%%%

In the CFL phase, all three condensates are nonzero. Here we obtain $\tilde{B}_3=\tilde{B}_8=0$ and Eqs.\ (\ref{eomrho}) yield three coupled equations for the condensates, which have the solution
\begin{subequations}\allowdisplaybreaks
\bea
\rho_1^2 &=& \frac{\lambda(\mu^2-m_1^2)+h(\mu^2+m_1^2-m_2^2-m_3^2)}{(\lambda-2h)(\lambda+h)} = \frac{\mu^2}{\lambda(1-2\eta)}\left(1-\frac{1}{3}\alpha\right) 
\, , \\[2ex]
\rho_2^2 &=& \frac{\lambda(\mu^2-m_2^2)+h(\mu^2-m_1^2+m_2^2-m_3^2)}{(\lambda-2h)(\lambda+h)} = \frac{\mu^2}{\lambda(1-2\eta)}\left[1-\frac{(7-2\eta)\alpha}{12(1+\eta)}\right] 
\, ,\\[2ex]
\rho_3^2 &=& \frac{\lambda(\mu^2-m_3^2)+h(\mu^2-m_1^2-m_2^2+m_3^2)}{(\lambda-2h)(\lambda+h)} = \frac{\mu^2}{\lambda(1-2\eta)}\left[1-\frac{(1+10\eta)\alpha}{12(1+\eta)}\right] \, , 
\eea
\end{subequations}
where, in the second steps, Eqs.\ (\ref{malpha}) and (\ref{eta}) have been used\footnote{Since the three condensates are different due to the strange quark mass, this phase was termed modified CFL (mCFL) in Ref.\ \cite{Iida:2004cj}. Here we simply keep the term CFL.}. In the massless limit, $\alpha=0$, we recover the result of three identical condensates.  The potential (\ref{U0}) becomes
\bea
U_{0}
&\simeq & -\frac{3\mu^4}{4\lambda(1-2\eta)} \left(1-\frac{2}{3}\alpha\right) \, ,
\eea
where we have dropped the  contribution quadratic in $\alpha$, i.e., all terms quartic in $m_s$.
This is consistent with our starting point, which only includes corrections to linear order in $\alpha$.
The Gibbs free energy density (\ref{gibbs}) thus becomes
\be \label{GCFL}
G_{\rm CFL} \simeq -\frac{3g^2H^2}{2(3g^2+4e^2)} -\frac{3\mu^4}{4\lambda(1-2\eta)} \left(1-\frac{2}{3}\alpha\right) \, ,
\ee
where we have used the explicit form of the mixing angles (\ref{thetamix}).

%%%%%%%%%%%%%%%%%%%%%%%%%%%%%%%%%%%%%%%%%%%%%%%%%%%%%%%%%%%%%%%
\subsection{2SC phases} 
\label{sec:2SC}
%%%%%%%%%%%%%%%%%%%%%%%%%%%%%%%%%%%%%%%%%%%%%%%%%%%%%%%%%%%%%%%

Next we discuss the three possible phases where exactly one condensate is non-vanishing. In each case, two flavors and two colors participate in pairing. 
According to the flavor structure of the condensates we term the phases 2SC$_{\rm ds}$, 2SC$_{\rm us}$, or 2SC$_{\rm ud}$ if $\rho_1$, $\rho_2$, or $\rho_3$ is 
non-vanishing, respectively. As we shall see, only the 2SC$_{\rm ud}$ phase is relevant for the phase structure since the other two phases turn out to be energetically disfavored. Therefore, we often use 2SC synonymously for 2SC$_{\rm ud}$. 

Starting with the most relevant phase, we first consider a nonzero $\rho_3$. 
From Eq.\ (\ref{eomrho3}) we then obtain 
\be \label{rho22SC}
\rho_3^2=\frac{\mu^2-m_3^2}{\lambda} = \frac{\mu^2}{\lambda}\left(1-\frac{\alpha}{12}\right) \, ,
\ee
and $\tilde{B}_8=0$. By inserting this into the Gibbs free energy density $G$ and minimizing with respect to 
$\tilde{B}_3$ we find 
\be \label{B32SC}
\tilde{B}_3= \frac{3egH}{\sqrt{3g^2+e^2}\sqrt{3g^2+4e^2}} \, , 
\ee
and inserting this back into $G$ yields 
\bea \label{Gud}
G_{{\rm 2SC}_{\rm ud}} &\simeq &  -\frac{3g^2H^2}{2(3g^2+e^2)}-\frac{\mu^4}{4\lambda}\left(1-\frac{\alpha}{6}\right)\, .
\eea
Analogously, if only the condensate $\rho_1$ is nonzero, i.e., for the 2SC$_{\rm ds}$ phase, we find 
\be
\rho_1^2 = \frac{\mu^2}{\lambda}\left(1-\frac{1}{3}\alpha\right)  \, ,
\ee
as well as $\tilde{B}_3=\tilde{B}_8=0$ from minimizing $G$, and we compute 
\be\label{Gds}
G_{{\rm 2SC}_{\rm ds}} \simeq  -\frac{3g^2H^2}{2(3g^2+4e^2)}  -\frac{\mu^4}{4\lambda}\left(1-\frac{2}{3}\alpha\right) \, .
\ee
Finally, for the 2SC$_{\rm us}$ phase, we find 
\be
\rho_2^2 = \frac{\mu^2}{\lambda}\left(1-\frac{7}{12}\alpha\right)   \, , 
\ee
and the magnetic fields 
\be
\tilde{B}_3=\frac{3eg(3g^2-2e^2)H}{2(3g^2+e^2)^{3/2}\sqrt{3g^2+4e^2}} \, , \qquad \tilde{B}_8=\frac{9eg^2H}{2(3g^2+e^2)^{3/2}} \, .
\ee
This yields the Gibbs free energy density 
\be\label{Gus}
G_{{\rm 2SC}_{\rm us}} \simeq  -\frac{3g^2H^2}{2(3g^2+e^2)} -\frac{\mu^4}{4\lambda}\left(1-\frac{7}{6}\alpha\right) \, .
\ee

Comparing Eqs.\ (\ref{Gud}), (\ref{Gds}), and (\ref{Gus}),  we see that for all values of the magnetic field $H$ and all nonzero values of $\alpha$, the 2SC$_{\rm ud}$ phase has the lowest Gibbs free energy and thus we can ignore the other two 2SC phases. In the massless limit $\alpha=0$ the 2SC$_{\rm ud}$ and 2SC$_{\rm us}$ phases become degenerate, as discussed in the introduction. 

%%%%%%%%%%%%%%%%%%%%%%%%%%%%%%%%%%%%%%%%%%%%%%%%%%%%%%%%%%%%%%
\subsection{fSC phases} 
\label{sec:dSC}
%%%%%%%%%%%%%%%%%%%%%%%%%%%%%%%%%%%%%%%%%%%%%%%%%%%%%%%%%%%%%%%

There are three possible phases in which exactly two condensates are non-vanishing. 
As for the 2SC phases, we shall see that one of these phases is favored over the other two for all parameter values. This is the phase where $\rho_2=0$.  In this case, there are $gd/bs$ and $gs/bd$ Cooper pairs from $\rho_1$ and  $ru/gd$ and $rd/gu$ Cooper pairs from $\rho_3$. Therefore, all three colors of the $d$ quark are involved in pairing (while only two colors of the $u$ and $s$ quarks are involved), and thus, following Ref.\ \cite{Iida:2004cj}, we refer to this phase as dSC. Analogously, the phases with vanishing $\rho_1$ and $\rho_3$ will be termed uSC and sSC, respectively. (Hence the collective term fSC, where f stands for the flavor.)

Starting again with the most relevant phase, we set $\rho_2=0$. With the help of Eqs.\ (\ref{eoma}) we obtain $\tilde{B}_3=\tilde{B}_8=0$. Then, Eqs.\ (\ref{eomrho1}) and (\ref{eomrho3}) yield two coupled equations for $\rho_1$ and $\rho_3$, with the solution 
\begin{subequations}
\bea
\rho_1^2 = \frac{\lambda(\mu^2-m_1^2)+h(\mu^2-m_3^2)}{\lambda^2-h^2} = \frac{\mu^2}{\lambda(1-\eta)}\left[1-\frac{(4+\eta)\alpha}{12(1+\eta)}\right]
\, ,\\[2ex]
\rho_3^2 = \frac{\lambda(\mu^2-m_3^2)+h(\mu^2-m_1^2)}{\lambda^2-h^2} =\frac{\mu^2}{\lambda(1-\eta)}\left[1-\frac{(1+4\eta)\alpha}{12(1+\eta)}\right] 
\, . 
\eea
\end{subequations}
Thus we compute 
\be \label{GdSC}
G_{\rm dSC} \simeq -\frac{3g^2H^2}{2(3g^2+4e^2)} -\frac{\mu^4}{2\lambda(1-\eta)}\left(1-\frac{5}{12}\alpha\right) \, ,
\ee
where again we have dropped the quadratic contribution in $\alpha$.

Analogously, if $\rho_3 =0$, we find again $\tilde{B}_3=\tilde{B}_8=0$, as well as 
\be
 \rho_1^2 = \frac{\mu^2}{\lambda(1-\eta)}\left[1-\frac{(4+7\eta)\alpha}{12(1+\eta)}\right]\, , \qquad
\rho_2^2 =\frac{\mu^2}{\lambda(1-\eta)}\left[1-\frac{(7+4\eta)\alpha}{12(1+\eta)}\right]
 \, , 
\ee
and
\be \label{GsSC}
G_{\rm sSC} \simeq -\frac{3g^2H^2}{2(3g^2+4e^2)} -\frac{\mu^4}{2\lambda(1-\eta)}\left(1-\frac{11}{12}\alpha\right) \, .
\ee
Finally, for $\rho_1=0$ we find $\tilde{B}_3=\tilde{B}_8=0$,
\be
\rho_2^2 =\frac{\mu^2}{\lambda(1-\eta)}\left[1-\frac{(7+\eta)\alpha}{12(1+\eta)}\right]  \, , \qquad \rho_3^2 = \frac{\mu^2}{\lambda(1-\eta)}\left[1-\frac{(1+7\eta)\alpha}{12(1+\eta)}\right]  \, , 
\ee
and 
\be \label{GuSC}
G_{\rm uSC} \simeq -\frac{3g^2H^2}{2(3g^2+4e^2)} -\frac{\mu^4}{2\lambda(1-\eta)}\left(1-\frac{2}{3}\alpha\right) \, .
\ee
Since the Gibbs free energy densities (\ref{GdSC}), (\ref{GsSC}), (\ref{GuSC}) of all  three phases receive the same contribution from the magnetic field, the one with the lowest energy penalty from the strange quark mass term is preferred (and they all become degenerate for vanishing mass). We can therefore focus on the dSC phase and ignore the other two phases. While in the massless case none of these phases appears in the phase diagram \cite{Haber:2017oqb}, we find that a window for this phase opens up in the presence of a strange quark mass, which was already observed (without external magnetic field) in Refs.\ \cite{Iida:2003cc,Iida:2004cj}.

%%%%%%%%%%%%%%%%%%%%%%%%%%%%%%%%%%%%%%%%%%%%%%%%%%%%%%%%%%%%%%%
\section{Critical magnetic fields}
\label{sec:HHH}
%%%%%%%%%%%%%%%%%%%%%%%%%%%%%%%%%%%%%%%%%%%%%%%%%%%%%%%%%%%%%%%

%%%%%%%%%%%%%%%%%%%%%%%%%%%%%%%%%%%%%%%%%%%%%%%%%%%%%%%%%%%%%%%
\subsection{Critical field $H_c$ }
\label{sec:Hc}
%%%%%%%%%%%%%%%%%%%%%%%%%%%%%%%%%%%%%%%%%%%%%%%%%%%%%%%%%%%%%%%

As we have just seen, the only relevant homogeneous phases within our ansatz 
are the CFL, 2SC$_{\rm ud}$, dSC, and NOR phases. From their Gibbs free energy densities we can now easily obtain the critical magnetic fields $H_c$ for the first-order transitions between them. In the usual terminology for ordinary superconductors, $H_c$ is the critical magnetic field for the first-order 
transition between the superconducting phase and the normal-conducting phase for a type-I superconductor. For a  color superconductor, the situation is more complicated because the various condensates can be broken sequentially until for sufficiently large magnetic field the normal phase is reached. As of now, we have not yet determined the transition from type-I to type-II behavior. To this end, the critical fields for the appearance of inhomogeneous phases have to be calculated, which we shall do in the subsequent sections.

By pairwise equating the Gibbs free energies (\ref{GNOR}), (\ref{GCFL}), (\ref{Gud}), and (\ref{GdSC}) we obtain the conditions for 6 potential phase transitions. There is one transition which is independent of the magnetic field, namely the CFL/dSC transition, which we can express for instance in terms of a critical value for $\eta$,
\be
\eta = -\frac{6-7\alpha}{2(3+\alpha)} \qquad \mbox{dSC/CFL} \, . 
\ee
Then, there are 4 transitions for which we can compute a critical magnetic field,  
\be \label{Hcs}
\frac{H_c^2}{\mu^4/\lambda} = \left\{\begin{array}{cc} \displaystyle{\left(1-\frac{1}{6}\alpha\right)\frac{3g^2+e^2}{2e^2}} & \mbox{2SC/NOR} \\[4ex] 
\displaystyle{\frac{3}{2(1-2\eta)}\left(1-\frac{2}{3}\alpha\right)\frac{3g^2+4e^2}{4e^2}} & \mbox{CFL/NOR} \\[4ex]
\displaystyle{\frac{1+\eta-(11+2\eta)\frac{\alpha}{12}}{1-2\eta}\frac{(3g^2+e^2)(3g^2+4e^2)}{9e^2g^2}} & \mbox{2SC/CFL} \\[4ex]
\displaystyle{\frac{1+\eta-(4+\eta)\frac{\alpha}{6}}{2(1-\eta)}\frac{(3g^2+e^2)(3g^2+4e^2)}{9e^2g^2}} & \mbox{2SC/dSC} \end{array}\right. \, .
\ee
The remaining transition, between the NOR and dSC phases formally has a critical field as well,
\be
\frac{H_c^2}{\mu^4/\lambda} = \frac{1-\frac{5}{12}\alpha}{1-\eta}\frac{3g^2+4e^2}{4e^2} \qquad \mbox{dSC/NOR} \, ,
\ee
but it turns out that this transition is never realized in the phase diagram. 

\begin{figure} [t]
\begin{center}
\hbox{\includegraphics[width=0.5\textwidth]{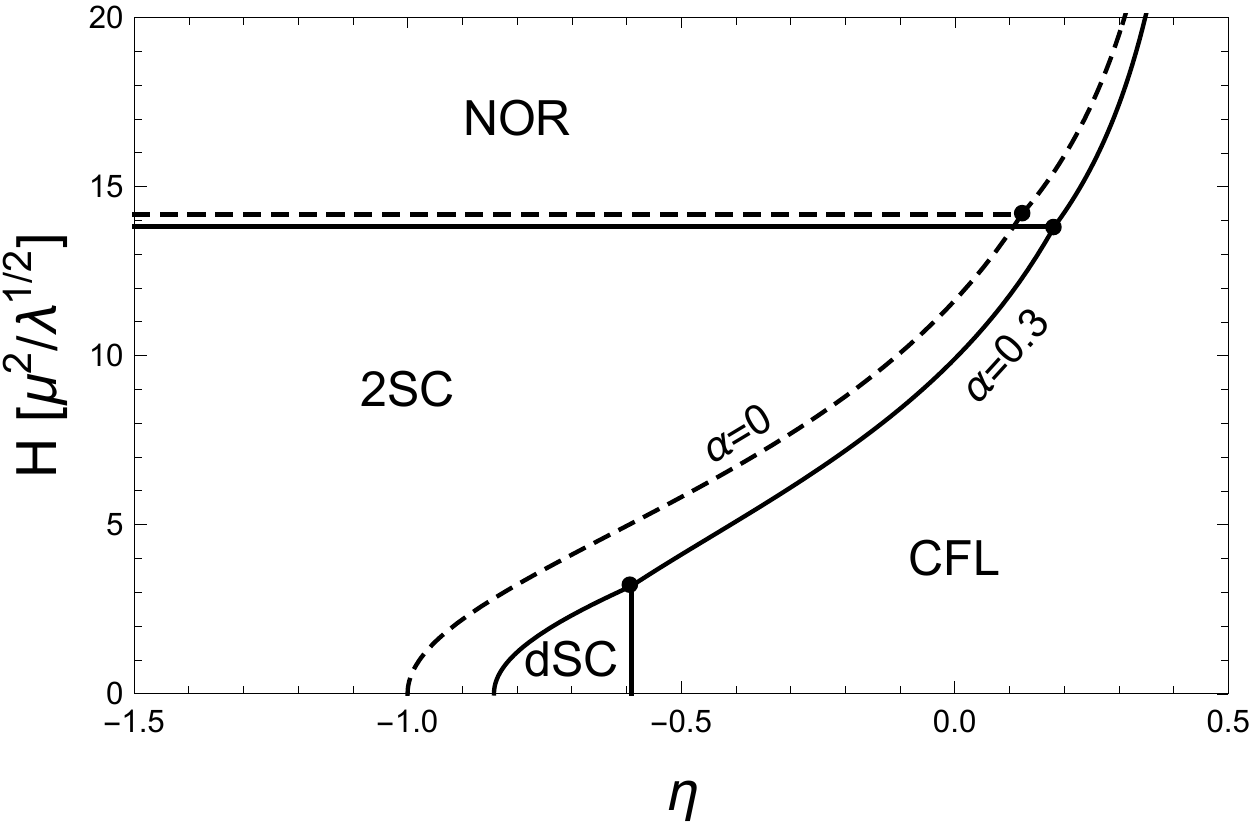}
\includegraphics[width=0.5\textwidth]{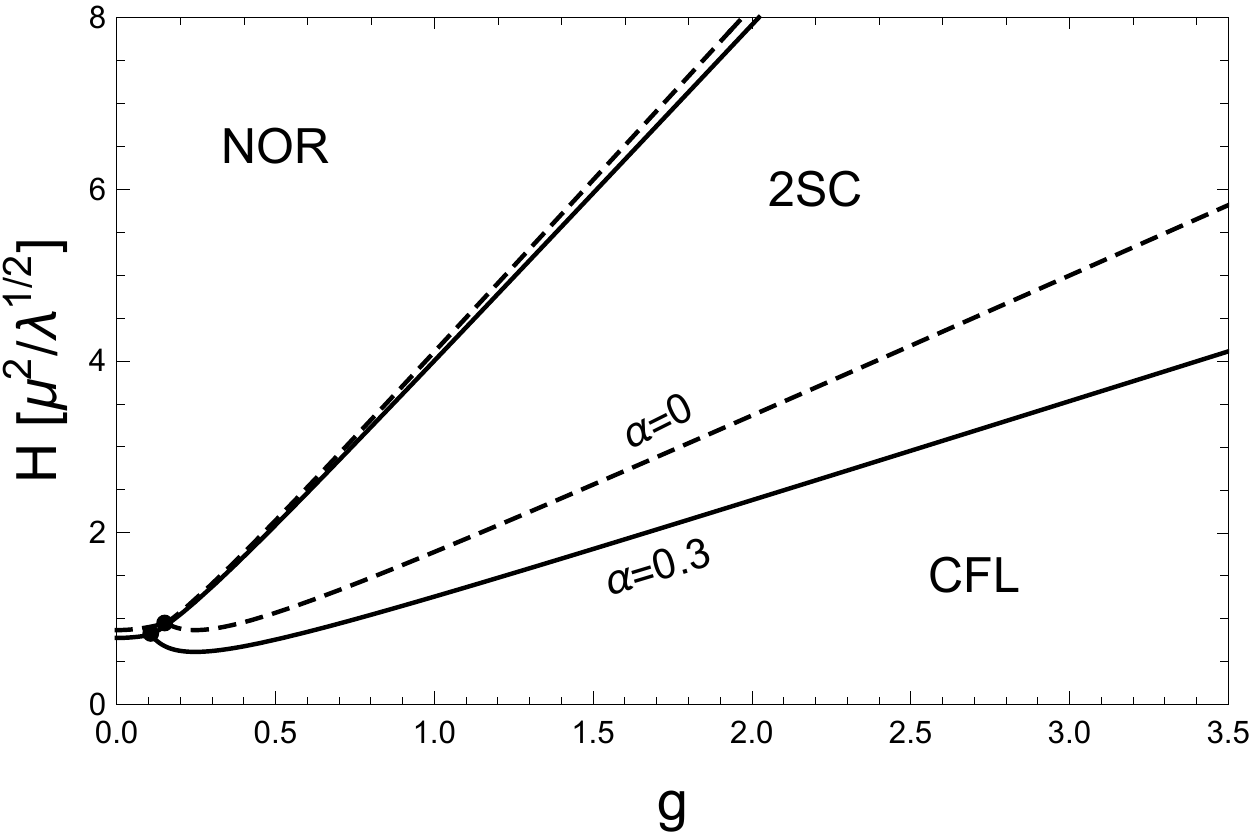}}
\caption{Homogeneous phases at nonzero external magnetic field $H$ in the plane spanned by $H$ and the parameter ratio  $\eta=h/\lambda$ (describing the cross-coupling between the condensates) for fixed strong coupling constant $g=3.5$ (left) and in the $H$-$g$ plane for fixed $\eta=-1/2$ (right). In both panels, the solid curves are for a mass parameter $\alpha=0.3$, while the dashed curves correspond to the massless limit, $\alpha=0$. 
}\label{fig:Heta}
\end{center}
\end{figure}

The phase structure must of course be invariant under the rotation chosen for the gauge fields. As a check, in the massless limit $\alpha=0$ one recovers the results of Ref.\ \cite{Haber:2017oqb}, where a different rotation was used.
We see that if the magnetic field is given in units of $\mu^2/\sqrt{\lambda}$, the parameter space has reduced to four dimensions, spanned 
by $H$, $\eta$, $\alpha$, and $g$ (for given $e\simeq 0.30$). In Fig.\ \ref{fig:Heta} we show two slices of this parameter space, namely in the $\eta$-$H$ plane for the value of $g$ used later, and in the $g$-$H$ plane for the value of $\eta$ used later. In each case we compare the massless result with the  $\alpha=0.3$ result. For a typical value $T_c/\mu_q \sim 0.1$ this value of $\alpha$ corresponds to $m_s/\mu_q\sim 0.5$. We see that for nonzero $\alpha$ there are two triple points, i.e., points where three phases have the same free energy. Both triple points are realized in the left panel. The one where CFL, 2SC, and NOR phases meet, and which also exists in the massless case, is given by
\be
\left(\eta,\frac{H}{\mu^2/\sqrt{\lambda}}\right) =\left(\frac{\frac{3}{8}\left(1+\frac{4}{3}\alpha\right)g^2-\left(1-\frac{11}{12}\alpha\right)e^2}{\left(1-\frac{\alpha}{6}\right)(3g^2+e^2)},\sqrt{\left(1-\frac{\alpha}{6}\right)\frac{3g^2+e^2}{2e^2}}\right) \, .
\ee
 This point also occurs in the right panel, where it has the coordinates
\bea
\left(g,\frac{H}{\mu^2/\sqrt{\lambda}}\right) &=& \left(\frac{2\sqrt{2}e}{\sqrt{3}}\sqrt{\frac{1-\frac{11}{12}\alpha+\eta\left(1-\frac{\alpha}{6}\right)}{1+\frac{4}{3}\alpha-8\eta\left(1-\frac{\alpha}{6}\right)}},\frac{3}{\sqrt{2}}\sqrt{\frac{1-\frac{5}{6}\alpha+\frac{\alpha^2}{9}}{1+\frac{4}{3}\alpha-8\eta\left(1-\frac{\alpha}{6}\right)}}\right) 
\, .
\eea
The other triple point, where dSC, 2SC, and CFL phases meet,  visible only in the left panel, is given  by 
\be \label{triple2}
\left(\eta,\frac{H}{\mu^2/\sqrt{\lambda}}\right) =\left(-\frac{6-7\alpha}{2(3+\alpha)},\frac{1}{6}\sqrt{\alpha\left(15+\frac{4e^2}{g^2}+\frac{9g^2}{e^2}\right)}\right) \, .
\ee
In the massless case, this point moves to the $H=0$ axis and is no longer a triple point because the intermediate dSC phase disappears in this limit.

To get an idea of the strength of the magnetic field in physical units, let us assume $\mu_q\simeq 400\, {\rm MeV}$, $T_c\simeq 0.05\mu_q$, which is typical for the interior of compact stars. Then, for instance, the critical magnetic field between the NOR and 2SC phases, $H_c \simeq 14\, \mu^2/\sqrt{\lambda}$, approximately corresponds to\footnote{Here we have used $1\, {\rm G}= 1\, {\rm g}^{1/2}{\rm cm}^{-1/2}{\rm s}^{-1}$ and thus  $1\, {\rm G} (c\hbar)^{3/2}=\beta\, {\rm eV}^2$ with the numerical factor $\beta\simeq 0.06925$. Together with $\sqrt{4\pi} \, B_{\rm HL} = B_{\rm G}$, where $B_{\rm HL}$ and $B_{\rm G}$ are the magnetic fields in the Heaviside-Lorentz and the Gaussian system of electromagnetism, we conclude that $1\, {\rm eV}^2$ in natural Heaviside-Lorentz units corresponds to $\sqrt{4\pi}/\beta\, {\rm G} \simeq 51.189\, {\rm G}$ in the Gaussian system.} $H_c\simeq 1.1\times 10^{19}\, {\rm G}$. 

The main effects of the strange quark mass, manifest in Fig.\ \ref{fig:Heta}, are as follows. As expected and observed in many other calculations, the strange quark mass tends to disfavor CFL compared to 2SC, and it also tends to slightly 
disfavor the 2SC phase compared to the normal phase. The dSC phase only appears in the presence of a strange quark mass, and does so already for vanishing magnetic field \cite{Iida:2004cj}. Both panels show that for the values $\eta=-1/2$ and $g=3.5$, which we shall use  in our calculation of the flux tubes, the dSC phase plays no role. As already pointed out in the massless case \cite{Haber:2017oqb}, there is a regime of small $g$ where the system transitions directly from the CFL phase to the NOR phase as we increase the magnetic field. Since our focus is on more realistic values of $g$ applicable to compact stars, we deal with the scenario where the transition to the NOR phase occurs via the 2SC phase. This is still true with a nonzero strange quark mass. 

Below a certain value of $\eta$, for instance $\eta\lesssim -0.85$ if $g=3.5$, the 2SC phase is the ground state even at vanishing magnetic field. This is worth mentioning since our main results concern 2SC flux tubes. For such a value of $\eta$, which goes beyond the weak-coupling
results, the 2SC phase is bounded by the NOR phase for large magnetic fields, but not bounded by any other phase for low magnetic fields, and thus the region in which 2SC flux tubes can be expected is  larger than for the weak-coupling value $\eta=-1/2$. In the weak-coupling scenario -- extrapolated to $g=3.5$ --  the 2SC phase is also bounded from below, namely by the CFL phase, and we shall see that this limits the region of 2SC flux tubes.

%%%%%%%%%%%%%%%%%%%%%%%%%%%%%%%%%%%%%%%%%%%%%%%%%%%%%%%%%%%%%%%
\subsection{Upper critical field $H_{c2}$}
\label{sec:Hc2}
%%%%%%%%%%%%%%%%%%%%%%%%%%%%%%%%%%%%%%%%%%%%%%%%%%%%%%%%%%%%%%%

Before we turn to the flux tubes themselves, it is useful to compute their upper 
critical field $H_{c2}$. In the standard scenario of a single condensate, this is the maximum magnetic field which can sustain a nonzero condensate, under the assumption of a second-order transition to the normal phase. It is therefore the  critical field below which an array of flux tubes is expected. In our multi-component system the situation is more complicated, and we have to calculate different critical fields $H_{c2}$ depending on which condensates melt. Having $H_c$ and $H_{c2}$ at hand, we can then determine the parameter regime where the color superconductor is of type II, and in particular where we expect 2SC flux tubes. 

 The calculation of $H_{c2}$ is a generalization to nonzero strange quark mass of the analogous calculation done in Ref.\ \cite{Haber:2017oqb}. That calculation, in turn, was a generalization of the standard single-component calculation which can be found in many textbooks. In a single-component superconductor, one linearizes the Ginzburg-Landau equations for a small condensate. The equation for the condensate then 
has the form of the Schr{\"o}dinger equation for the harmonic  oscillator, from which one reads off the maximal possible magnetic field $H_{c2}$, corresponding to the ground state energy. We know from the previous subsection that for strong coupling, as we decrease the magnetic field within the NOR phase, we encounter the 2SC phase. Consequently, for the corresponding critical field $H_{c2}$ we only have to take into account a single condensate, i.e., this case is analogous to the textbook scenario and leads to the simple generalization of Eq.\ (40) in Ref.\ \cite{Haber:2017oqb},
\be \label{Hc22SCNOR}
\mbox{2SC/NOR:} \qquad H_{c2} = \frac{3(\mu^2-m_3^2)}{e} = \frac{3\mu^2}{e}\left(1-\frac{\alpha}{12}\right) \, .
\ee
In this standard scenario all three critical magnetic fields $H_c$, $H_{c1}$, and $H_{c2}$ intersect at a single point (as a function of a model parameter, usually the Ginzburg-Landau parameter $\kappa$). Therefore, this intercept defines the transition between type-I and type-II behavior (usually at $\kappa=1/\sqrt{2}$). 
Here, by equating $H_{c2}$ with $H_c$ for the 2SC/NOR transition in Eq.\ (\ref{Hcs}) we find for this transition point 
\be \label{type2SCNOR}
\frac{T_c}{\mu_q} = \frac{\sqrt{7\zeta(3)}}{12\sqrt{3}\pi^2} \sqrt{g^2+\frac{e^2}{3}} +{\cal O}\left(\frac{m_s^4}{\mu_q^4}\right) \, ,
\ee
where the weak-coupling expression of $\lambda$ in Eq.\ (\ref{weak}) has been used. It thus turns out that $T_c/\mu_q$ is a natural parameter to distinguish between type-I and type-II behavior -- with large $T_c/\mu_q$ corresponding to type II. Interestingly, we see that there is no mass correction to the transition point within the order of our approximation. We shall see later that indeed $H_{c1}$ intersects $H_c$ and $H_{c2}$ at the same $T_c/\mu_q$. The reason is that in the vicinity of this point the system effectively behaves as a single-component system. Additional condensates can be induced  in the cores of 2SC flux tubes -- and our main results concern such unconventional flux tubes -- but we will see that this is not the case close to the point (\ref{type2SCNOR}). 

The transition from the homogeneous 2SC phase, where $\phi_3$ is nonzero,  to an inhomogeneous phase is slightly more complicated.  Assuming a second-order transition, we linearize the Ginzburg-Landau equations in $\phi_1$ and $\phi_2$. In the massless limit, the resulting two (decoupled) equations yield the same critical magnetic field \cite{Haber:2017oqb}. In other words, as we approach the flux tube phase by decreasing $H$, both $\phi_1$ and $\phi_2$ become nonzero simultaneously (and continuously). This is different for nonzero $m_s$, in which case the two relevant equations are 
\begin{subequations}
\bea
\left[\left(\nabla+i\tilde{q}_3
\tilde{\bf A}_3\right)^2+(\mu^2-m_1^2)+2h|\phi_3|^2\right]\phi_1 &\simeq& 0 \, , \\[2ex]
\left[\left(\nabla-i\tilde{q}_3
\tilde{\bf A}_3\right)^2+(\mu^2-m_2^2)+2h|\phi_3|^2\right]\phi_2 &\simeq& 0 \, ,
\eea
\end{subequations}
where we have set $\tilde{\bf A}_8=0$ since $\tilde{B}_8=0$ in the 2SC phase and where $\phi_3 = \rho_3/\sqrt{2}$ is the condensate in the homogeneous 2SC phase (\ref{rho22SC}).   With the usual arguments and using the 2SC relation between $\tilde{B}_3$ and $H$ from Eq.\ (\ref{B32SC}), we obtain two different critical fields,
\begin{subequations} \label{Hc212}
\bea
H_{c2}^{(1)} &=& \frac{2(3g^2+e^2)}{3eg^2}[\mu^2(1+\eta)-(m_1^2+\eta m_3^2)] = \frac{2\mu^2(3g^2+e^2)}{3eg^2}\left(1+\eta-\frac{4+\eta}{12}\alpha\right)  \, , \label{Hc2one}
\\[2ex]
H_{c2}^{(2)} &=& \frac{2(3g^2+e^2)}{3eg^2}[\mu^2(1+\eta)-(m_2^2+\eta m_3^2)] = \frac{2\mu^2(3g^2+e^2)}{3eg^2}\left(1+\eta-\frac{7+\eta}{12}\alpha\right) \, . 
\eea
\end{subequations} 
The most relevant case for us is the one where both $H_{c2}^{(1)}$  and $H_{c2}^{(2)}$ are positive (a formally negative value indicates that the critical field does not exist, indicating that the homogeneous phase persists down to $H=0$). This is the case for $\eta=-1/2$ and all reasonable, i.e., not too large, values of $\alpha$. In this scenario, there is a transition at  $H_{c2}^{(1)}$ from the homogeneous 2SC phase to a phase where both $\phi_1$ and $\phi_3$ are nonzero, which is an inhomogeneous version of the dSC phase (see Sec.\ \ref{sec:dSC}). Then, as we reach the "would-be" $H_{c2}^{(2)}$ by further decreasing $H$, the approximation by which this critical field was computed is no longer valid, and thus the value for $H_{c2}^{(2)}$ becomes irrelevant. Nevertheless, it can be expected that there will be some transition from an inhomogeneous dSC phase to an inhomogeneous CFL phase. The existence of an intermediate inhomogeneous dSC phase due to  the nonzero strange quark mass is an interesting new observation, but it is beyond the scope of this paper to construct this phase explicitly. 

We may again compute the transition point between type-I and type-II behavior. For $\eta=-1/2$ (where there is no homogeneous dSC phase), we equate $H_{c2}^{(1)}$ with $H_c$ for the 2SC/CFL transition from Eq.\ (\ref{Hcs}). Dropping terms quadratic in $\alpha$ we find
\be
\frac{T_c}{\mu_q} \simeq  c\left(1-\frac{\alpha}{4}\right) \, , \qquad 
c\equiv \frac{\sqrt{7\zeta(3)}}{12\sqrt{2}\pi^2}\frac{g\sqrt{3g^2+4e^2}}{\sqrt{3g^2+e^2}} \,.
\ee
Since $\alpha$ depends on $T_c/\mu_q$ this is an implicit equation for $T_c/\mu_q$. To lowest nontrivial order in $m_s^2/\mu_q^2$ the solution is 
\be \label{type2SCCFL}
\frac{T_c}{\mu_q} = c\left(1+\frac{m_s^2}{8\mu_q^2}\ln c\right) +{\cal O}\left(\frac{m_s^4}{\mu_q^4}\right) \, .
\ee
 Therefore, this transition point between type-I and type-II behavior {\it does} receive a correction quadratic in $m_s$ (linear in $\alpha$), in contrast to the transition point  (\ref{type2SCNOR}).  The detailed phase structure around this point is expected to be complicated. This is due to the intermediate inhomogeneous dSC phase, as just discussed, but even without mass correction this transition point is affected in a nontrivial way by the multi-component nature of the system \cite{Haber:2017kth,Haber:2017oqb}. Most importantly, if the lower boundary of the flux tube region $H_{c1}$ is computed in the usual way, i.e., assuming a second-order transition, it turns out that the three critical fields no longer intersect in a single point, and the situation becomes more complicated due to a first-order entrance into the flux tube phase. Here we do not have to deal with these complications, since the precise location of the transition point (\ref{type2SCCFL}) and the phase transitions in its vicinity are not relevant for the 2SC flux tubes.

%%%%%%%%%%%%%%%%%%%%%%%%%%%%%%%%%%%%%%%%%%%%%%%%%%%%%%%%%%%%%%%
\subsection{Flux tubes and lower critical field $H_{c1}$}
\label{sec:Hc1}
%%%%%%%%%%%%%%%%%%%%%%%%%%%%%%%%%%%%%%%%%%%%%%%%%%%%%%%%%%%%%%%

Having identified the parameter range where type-II behavior with respect to 2SC flux tubes is expected, we can now turn to the explicit construction of these flux tubes. We will restrict ourselves to the calculation of an isolated, straight flux tube, such that we can employ cylindrical symmetry and our calculation becomes effectively one-dimensional in the radial direction. This is sufficient to compute the critical field $H_{c1}$, which is defined as the  field at which it becomes favorable to place a single flux tube in the system, indicating a second-order transition to a phase containing an array of flux tubes. Since the distance between the flux tubes goes to infinity as $H_{c1}$ is approached from above, the interaction between flux tubes plays no role. As explained in the introduction, our main goal is to determine the fate of the 2SC domain walls in the presence of a nonzero strange quark mass. Therefore, we focus exclusively on 2SC flux tubes, i.e., configurations which asymptote to the 2SC phase far away from the center of the flux tube. 

In order to compute the profiles of the condensates and the gauge fields we need to derive their equations 
of motion and bring them into a form convenient for the numerical evaluation. We work in cylindrical coordinates $(r,\theta,z)$, where, as above, the $z$-axis is aligned with the external magnetic field ${\bf H}$
and thus with the flux tube. We introduce dimensionless condensates $f_i$ ($i=1,2,3$),  which only depend on the radial distance to the center of the flux tube, 
\be
\rho_i({\bf r}) = f_i(r)\rho_{\rm 2SC} \, , 
\ee
where we have denoted the condensate of the homogeneous 2SC phase (\ref{rho22SC}) by $\rho_{\rm 2SC}$. Since we are interested in 2SC flux tubes, we impose the boundary conditions $f_3(\infty)=1$ and $f_1(\infty)=f_2(\infty)=0$. As in ordinary single-component flux tubes, we allow for a nonzero winding number $n\in \mathbb{Z}$, such that the phases of the condensates are 
\be \label{windings}
\psi_1({\bf r})=\psi_2({\bf r})=0 \, , \qquad \psi_3({\bf r})=n\theta \, .
\ee
Here we have set the winding numbers for the "non-2SC" condensates $f_1$ and $f_2$ to zero. In principle, we might include configurations where these windings are nonzero. (The baryon circulation around the flux tube vanishes for arbitrary 
choices of the winding numbers as long as $f_1(\infty)=f_2(\infty)=0$.) In such configurations, $f_1$ and/or $f_2$ would have to vanish far away from the flux tube {\it and} in the center of the flux tube, i.e., at best they would be non-vanishing in an intermediate domain. These configurations do not play a role in the massless limit \cite{Haber:2017oqb} and there is no obvious reason why they should become important if a strange quark mass is taken into account. Therefore, we shall work with Eq.\ (\ref{windings}). As a consequence, the boundary condition for the 2SC condensate in the core is $f_3(0)=0$, while $f_1(0)$ and $f_2(0)$ can be nonzero and must be determined dynamically. 

As we have seen in Sec.\ \ref{sec:setup}, after the rotation of the gauge fields $\tilde{\bf A}$ decouples from the condensates. Therefore, we are left with two nontrivial gauge fields, for which we introduce the dimensionless versions $\tilde{a}_3$ and $\tilde{a}_8$ via
\be \label{A3A8tt}
 \tilde{\bf A}_3({\bf r}) = \left[\frac{H\cos\vartheta_1\sin\vartheta_2}{2}r+\frac{\tilde{a}_3(r)}{r}\right]{\bf e}_\theta \, ,
 \qquad \tilde{\bf A}_8({\bf r}) = \frac{\tilde{a}_8(r)}{r}{\bf e}_\theta \, ,
\ee
with the boundary conditions $\tilde{a}_3(0) = \tilde{a}_8(0)=0$. This yields the magnetic fields 
\be \label{B3B8}
\tilde{\bf B}_3 = \left(H\cos\vartheta_1\sin\vartheta_2+\lambda\rho_{\rm 2SC}^2\frac{\tilde{a}_3'}{R}\right){\bf e}_z \, , \qquad \tilde{\bf B}_8 =\lambda\rho_{\rm 2SC}^2\frac{\tilde{a}_8'}{R}
{\bf e}_z \, ,
\ee
where prime denotes derivative with respect to $R$, which is the dimensionless radial coordinate
\be \label{R}
R=\frac{r}{\xi_3} \, , \qquad \xi_3 \equiv \frac{1}{\sqrt{\lambda}\rho_{\rm 2SC}} \, . 
\ee
Here, $\xi_3$ is the coherence length; we have added a subscript 3 to indicate that in a 2SC flux tube it is the $ud$ condensate $f_3$ whose asymptotic behavior is characterized by the coherence length. Following Ref.\ \cite{Haber:2017oqb}, we have separated a term in $\tilde{\bf A}_3$ for convenience, which gives rise to the nonzero field $\tilde{\bf B}_3$ in the 2SC phase, i.e., far away from the flux tube. Therefore, the dimensionless gauge fields do not create any additional magnetic fields at infinity, $\tilde{a}_3'(\infty)=\tilde{a}'_8(\infty)=0$ (alternatively, this effect could have been implemented in the boundary condition for $\tilde{a}_3$). The behavior of $\tilde{B}_3$ is another qualitative difference of the 2SC flux tube to a textbook flux tube (besides potentially induced additional condensates and the existence of two gauge fields). In a standard flux tube, the magnetic field is expelled in the superconducting phase far away from the flux tube and penetrates through the normal-conducting centre. Here, we have three magnetic fields: $\tilde{B}$, which fully penetrates the superconductor and thus is irrelevant for the calculation of the flux tube profiles; $\tilde{B}_8$, which behaves analogously to the ordinary magnetic field in an ordinary flux tube; and $\tilde{B}_3$, which is nonzero far away from the flux tube {\it and} is affected nontrivially by the flux tube profile. As a consequence, since $\tilde{B}_3$ depends on the external field $H$, the flux tube profiles and flux tube energies also depend on $H$, which poses a technical complication. We should keep in mind, however, that it is the ordinary magnetic field ${\bf B}$ that dictates the formation of magnetic defects. A flux tube configuration, in which the condensation energy is necessarily reduced, may become favored if this energy cost is overcompensated by admitting magnetic ${\bf B}$-flux in the system (this is the meaning of the $-{\bf B}\cdot{\bf H}$ term in the Gibbs free energy). It is therefore useful for the interpretation of our results to compute the unrotated field ${\bf B}$ from the profiles. Undoing the rotation (\ref{rotatetwice}) and using Eqs.\ (\ref{tildeB}) and (\ref{B3B8}), we find 
\be \label{BoverH}
\frac{B}{H} = \cos^2\vartheta_1\left[1 +\frac{e\sin\vartheta_1}{4\Xi}\left( \frac{\tilde{a}_8'}{R} +\frac{3g}{\sqrt{3g^2+4e^2}}\frac{\tilde{a}_3'}{R}\right)\right] \, ,
\ee
where we have introduced the  dimensionless magnetic field 
\be \label{Xi}
\Xi \equiv \frac{\tilde{q}_3 H\cos\vartheta_1\sin\vartheta_2}{2\lambda\rho_{\rm 2SC}^2} = \frac{3eg^2}{4\sqrt{\lambda}(3g^2+e^2)}\left(1-\frac{m_3^2}{\mu^2}\right)^{-1}\frac{H}{\mu^2/\sqrt{\lambda}} \, .
\ee 
We can now express the potential $U_0$ (\ref{U0}) in terms of the dimensionless condensates and gauge fields,
\bea\label{U01}
U_0  &=& U_{\rm 2SC} + \frac{\lambda\rho_{\rm 2SC}^4}{2}\left[f_1'^2+f_2'^2+f_3'^2+f_1^2\left(\frac{f_1^2}{2}-\frac{\mu^2-m_1^2}{\mu^2-m_3^2}\right) +f_2^2\left(\frac{f_2^2}{2}-\frac{\mu^2-m_2^2}{\mu^2-m_3^2}\right)+\frac{(1-f_3^2)^2}{2} \right.\non[2ex]
&&\left. +\frac{({\cal N}_1+\Xi R^2)^2f_1^2+({\cal N}_2-\Xi R^2)^2f_2^2+{\cal N}_3^2f_3^2}{R^2} -\eta(f_1^2f_2^2+f_1^2f_3^2+f_2^2f_3^2)\right] \, .
\eea
Here we have denoted the potential of the homogeneous 2SC phase 
by 
\be
U_{\rm 2SC}=-\frac{(\mu^2-m_3^2)^2}{4\lambda}\, , 
\ee
and we have abbreviated 
\bea
{\cal N}_1 &\equiv& \tilde{q}_3\tilde{a}_3+\tilde{q}_{81}\tilde{a}_8 \, ,\qquad 
{\cal N}_2 \equiv -\tilde{q}_3\tilde{a}_3+\tilde{q}_{82}\tilde{a}_8 \, ,\qquad 
{\cal N}_3 \equiv n-\tilde{q}_{83}\tilde{a}_8 \, . 
\eea
Inserting Eq.\ (\ref{U01}) into the Gibbs free energy density (\ref{gibbs}) and using the expressions for the magnetic fields (\ref{B3B8}), we derive the equations of motion for the gauge fields,
\begin{subequations} \label{eqsa}
\bea
\tilde{a}_3''-\frac{\tilde{a}_3'}{R}&=& \frac{\tilde{q}_3}{\lambda}[({\cal N}_1+\Xi R^2)f_1^2-({\cal N}_2-\Xi R^2)f_2^2] \, ,\label{a3t}  \\[2ex]
\tilde{a}_8''-\frac{\tilde{a}_8'}{R}&=& \frac{1}{\lambda}[\tilde{q}_{81}({\cal N}_1+\Xi R^2)f_1^2+\tilde{q}_{82}({\cal N}_2-\Xi R^2)f_2^2-\tilde{q}_{83}{\cal N}_3f_3^2] \, , \label{a8t}
\eea
\end{subequations}
and for the condensates, 
\begin{subequations} \label{eqsf}
\bea
0&=& f_1''+\frac{f_1'}{R}+f_1\left[\frac{\mu^2-m_1^2}{\mu^2-m_3^2}-f_1^2-\frac{({\cal N}_1+\Xi R^2)^2}{R^2} +\eta(f_2^2+f_3^2)\right] \, , \\[2ex]
0&=& f_2''+\frac{f_2'}{R}+f_2\left[\frac{\mu^2-m_2^2}{\mu^2-m_3^2}-f_2^2-\frac{({\cal N}_2-\Xi R^2)^2}{R^2} +\eta(f_1^2+f_3^2)\right] \, , \label{eqf2} \\[2ex]
0&=& f_3''+\frac{f_3'}{R}+f_3\left[1-f_3^2-\frac{{\cal N}_3^2}{R^2} +\eta(f_1^2+f_2^2)\right] \, . \label{eqf3}
\eea
\end{subequations}
By taking the limit $R\to \infty$ of Eq.\ (\ref{a8t})  we conclude
\be \label{a8infty}
\tilde{a}_8(\infty) = \frac{n}{\tilde{q}_{83}} \, ,
\ee
while no condition for $\tilde{a}_3(\infty)$ can be derived, hence this value has to be determined dynamically. 
Due to the boundary value (\ref{a8infty}), the baryon circulation around the flux tube vanishes, as in a standard magnetic flux tube. We discuss the asymptotic behavior far away from the center of the flux tube in Appendix \ref{appA}.

The Gibbs free energy density can be written as  
\bea \label{Gflux}
G 
&=&U_{\rm 2SC}-\frac{H^2\cos^2\vartheta_1}{2}+\frac{L}{V} \left({\cal F}-\tilde{\Phi}_8H\sin\vartheta_1\right) \, , 
\eea
where $L$ is the size of the system in the $z$-direction, and 
\be
\tilde{\Phi}_8 = \oint d{\bf s}\cdot\tilde{\bf A}_8 = \frac{2\pi n}{\tilde{q}_{83}} \, ,  
\ee
 with a closed integration contour encircling the flux tube at infinity, is the magnetic $\tilde{\bf B}_8$-flux through the flux tube. Employing  partial integration and the  equations of motion (\ref{eqsf}), the free energy of a single flux tube per unit length is  
${\cal F} = \pi\rho_{\rm 2SC}^2{\cal I}$ with 
\be
{\cal I} \equiv \int_0^\infty dR\,R\left[\frac{\lambda(\tilde{a}_3'^2+\tilde{a}_8'^2)}{R^2}-\frac{f_1^4}{2}-\frac{f_2^4}{2}+\frac{1-f_3^4}{2}+\eta
(f_1^2f_2^2+f_1^2f_3^2+f_2^2f_3^2)\right] \, .
\ee
 Written in this form, the free energy does not have any explicit dependence on the mass correction and is identical to the one in Ref.\ \cite{Haber:2017oqb} (of course, the dependence on the mass enters implicitly through the equations
of motion). 
The critical field $H_{c1}$ is defined as the field above which $G$ is lowered by the addition of a flux tube, i.e., by the point at which the term in parentheses in Eq.\ (\ref{Gflux}) is zero.  This condition can be written as 
\be \label{Xictube}
\Xi_{c1} =\frac{g^2 {\cal I}(\Xi_{c1},n)}{8\lambda n}  \, ,
\ee
where $\Xi_{c1}$ is the dimensionless version of $H_{c1}$  via Eq.\ (\ref{Xi}). 
In the ordinary textbook scenario, the free energy of a flux tube does not depend on the external magnetic field,  and thus  (\ref{Xictube}) would be an explicit expression for the (dimensionless) critical magnetic field. The free energy of a 2SC flux tube, however, {\it does} depend on the external magnetic field. Therefore, Eq.\ (\ref{Xictube})
is an {\it implicit} equation for $\Xi_{c1}$, which has to be solved numerically. 

As we shall see in the next section, $H_{c1}$ may intersect with $H_{c2}^{(1)}$, which indicates the second-order transition from the homogeneous 2SC phase to an inhomogeneous phase where the condensate $f_1$ is switched on. Since $H_{c1}$ becomes meaningless beyond this intercept, it is useful to compute this point explicitly. To this end, we write $H_{c2}^{(1)}$ from Eq.\ (\ref{Hc2one}) (setting $\eta=-1/2$) in terms of the dimensionless field $\Xi$ (\ref{Xi}), which yields to linear order in $\alpha$ 
\be \label{Xic2}
\Xi_{c2}^{(1)} \simeq  \frac{1}{4}\left(1-\frac{\alpha}{2}\right)\, .
\ee
Solving this equation simultaneously with Eq.\ (\ref{Xictube}), i.e., setting $\Xi_{c2}^{(1)} =\Xi_{c1}$, gives the intersection point. In the practical calculation, this is best done by inserting Eq.\ (\ref{Xic2}) into Eq.\ (\ref{Xictube}) and solving the resulting equation for $T_c/\mu_q$ (for given strange quark mass and winding number).
This calculation is relevant for the phase diagrams in Fig.\ \ref{fig:phasediagram}.

%%%%%%%%%%%%%%%%%%%%%%%%%%%%%%%%%%%%%%%%%%%%%%%%%%%%%%%%%%%%%%%
\section{Numerical results and discussion}
\label{sec:results}
%%%%%%%%%%%%%%%%%%%%%%%%%%%%%%%%%%%%%%%%%%%%%%%%%%%%%%%%%%%%%%%

We are now prepared to compute the flux tube profiles and the resulting critical fields $H_{c1}$, which we will put together with $H_c$ and $H_{c2}$ from the previous section. As in the calculations of the critical fields $H_c$ and $H_{c2}$ we use the bosonic masses (\ref{m123}), keep terms linear in $\alpha$ in Eqs.\ (\ref{eqsf}), and, as discussed in Sec.\ \ref{sec:para}, we set $\eta=-1/2$ and $g=3.5$ for all following results. Then we solve the coupled second-order differential equations (\ref{eqsa}) and (\ref{eqsf}) numerically without further approximations using the successive over-relaxation method, which has been used before in similar contexts \cite{Haber:2017kth,Haber:2017oqb,Haber:2018tqw,Haber:2018yyd,Fraga:2018cvr,Schmitt:2020tac}.  Each numerical run yields a flux tube profile for given $m_s/\mu_q$, $T_c/\mu_q$ (from which $\alpha$ and $\lambda$ are obtained), dimensionless external magnetic field $\Xi$, and winding number $n$. Since we are interested in the critical field $\Xi_{c1}$, we need to solve Eq.\ (\ref{Xictube}), for which we employ the bisection method. This requires solving the differential equations about 10 -- 20 times until a reasonable accuracy for $\Xi_{c1}$ is reached. This whole procedure thus yields a critical field $H_{c1}$ for given $m_s/\mu_q$, $T_c/\mu_q$, and $n$. As argued in the introduction and as we shall see below, solutions with high winding numbers are expected to play an important role. Therefore, 
in principle, the procedure has to be repeated for all $n$ to find the preferred flux tube configuration for each point in the parameter space spanned by $m_s/\mu_q$ and $T_c/\mu_q$. In practice, we have performed the calculation for the lowest few $n$, and for selected parameter sets for much larger $n$ to check our conclusions.
An additional complication arises because in certain parameter regions there is more than one solution to the set of differential equations. The single-component flux tube with $f_1\equiv f_2\equiv 0$ always exists. Configurations where the condensate $f_2$ is induced in the core of the flux tube, but not $f_1$, turn out to be preferred over the single-component configuration whenever they exist and we shall discuss them in detail. Configurations where both $f_1$ and $f_2$ are induced in the core do exist as well, but only in a parameter region where $H_{c2}$ indicates that the ground state is a more complicated flux tube array. We shall therefore not discuss these three-component configurations.

%%%%%%%%%%%%%%%%%%%%%%%%%%%%%%%%%%%%%%%%%%%%%%%%%%%%%%%%%%%%%%%
\subsection{Flux tube properties}
\label{sec:tubes}
%%%%%%%%%%%%%%%%%%%%%%%%%%%%%%%%%%%%%%%%%%%%%%%%%%%%%%%%%%%%%%%

\begin{figure} [h]
\begin{center}
\includegraphics[width=\textwidth]{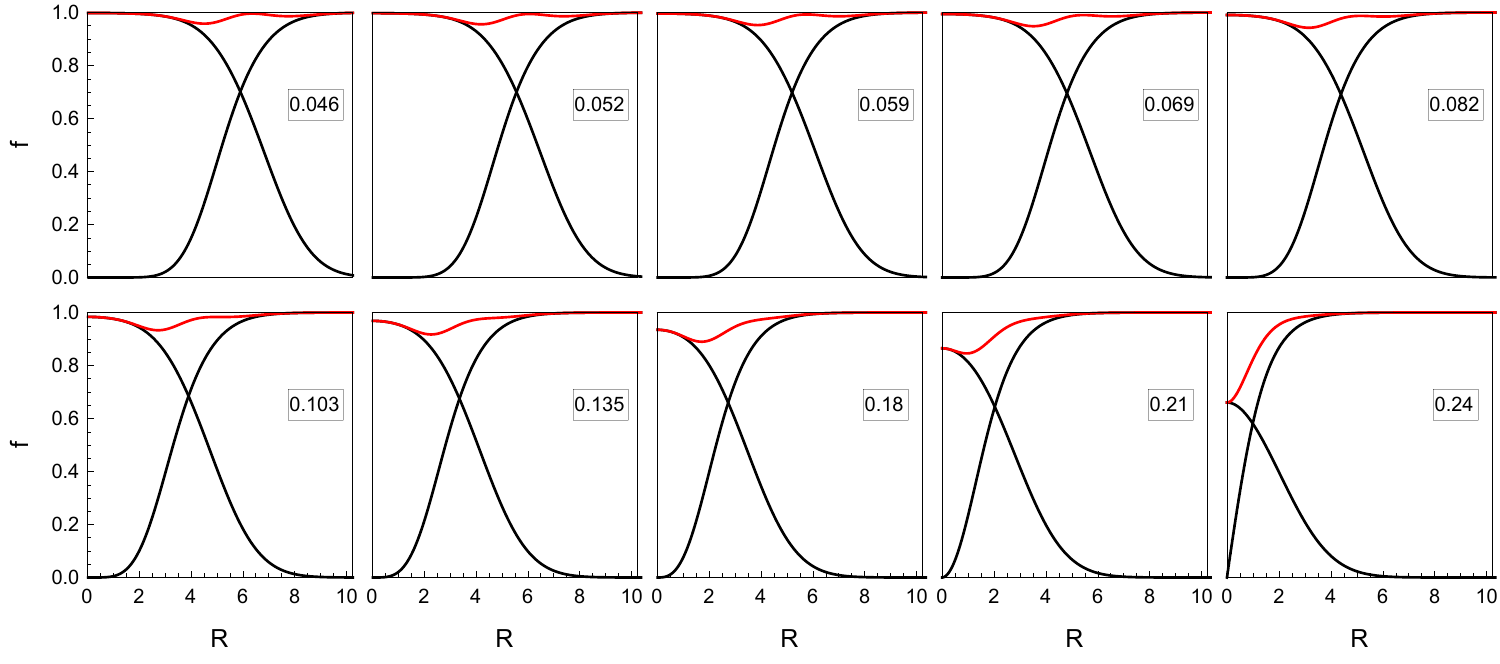}

\vspace{0.5cm}
\includegraphics[width=\textwidth]{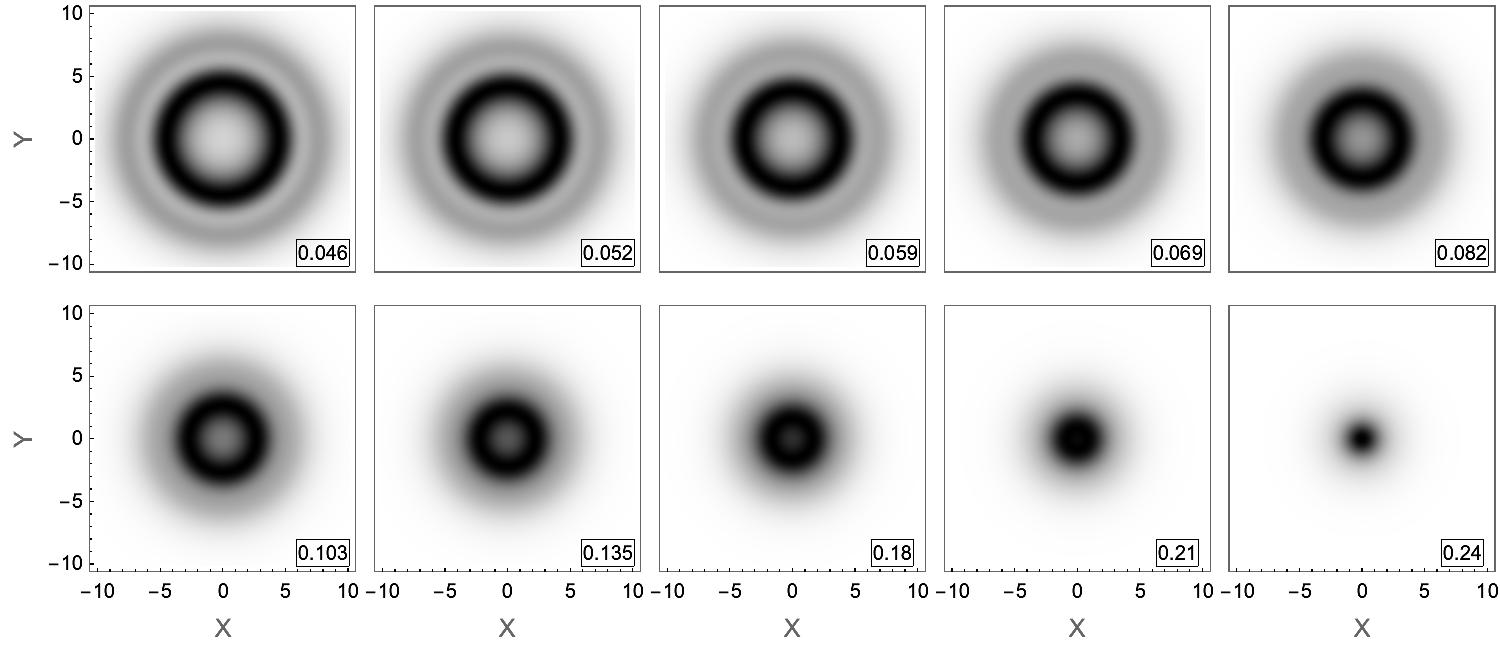}
\caption{{\it Upper panels:} Flux tube profiles of the dimensionless condensates $f_2$ and $f_3$ (black) and $\sqrt{f_2^2+f_3^2}$ (red) as  functions of the dimensionless radial coordinate $R$. The condensate $f_3$ is the usual 2SC condensate of $ud$ Cooper pairs and asymptotes to 1 for large $R$, while the $us$ condensate $f_2$ is induced in the core of the flux tube. For all plots $g=3.5$ and $T_c=0.0856\mu_q$, while $m_s/\mu_q$ assumes the values given in each panel, i.e., it increases from upper left to lower right. The masses are chosen such that the 
preferred configurations are flux tubes with winding numbers $n=10,\ldots, 1$ from upper left to lower right (and each plot shows the preferred configuration). For compact star conditions, $R=10$ translates to about $r\simeq 7.7\, {\rm fm}$. {\it Lower panels:} Ratio of the induced magnetic field over the 
external magnetic field, $B/H$, in the $X$-$Y$ plane perpendicular to the flux tube for the 10 configurations of the upper panels ($X$ and $Y$ in the same dimensionless units as $R$). The scale of the shading is adjusted for each plot separately, from black (maximal) to white (minimal). The magnetic field enters the superconductor in ring-like structures for small strange quark mass (large winding number) and turns into the conventional flux tube behavior for large mass (small winding number). This structure is reflected in the red curves of the upper panels. 
}\label{fig:profiles}
\end{center}

\vspace{-0.8cm}
\end{figure}

\begin{figure} [t]
\begin{center}
\includegraphics[width=0.5\textwidth]{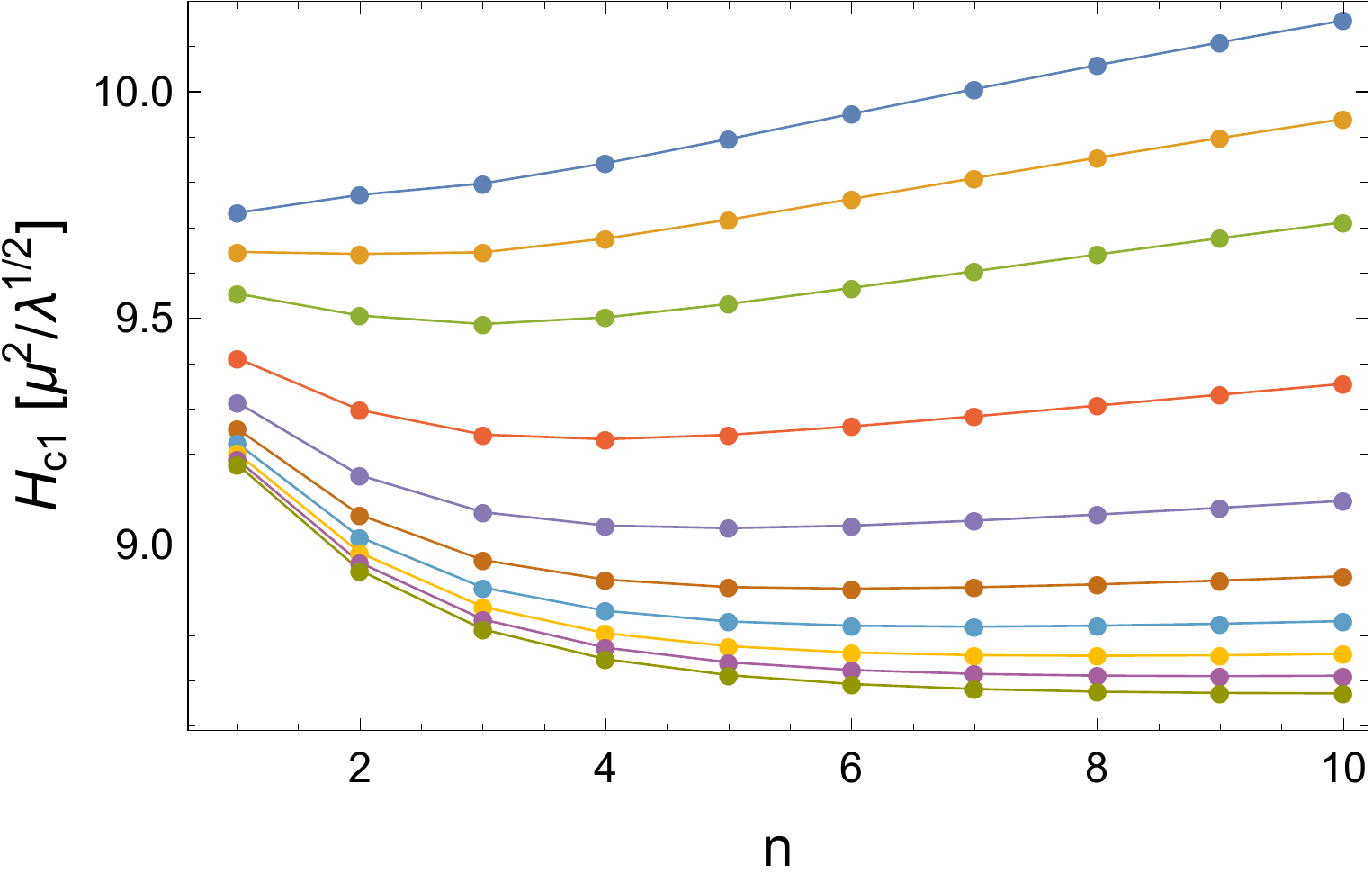}
\caption{Critical magnetic fields $H_{c1}$ as a function of the winding number $n$ (continuous lines inserted to guide the eye) for the same parameters as in Fig.\ \ref{fig:profiles}. Each line corresponds to a different strange quark mass, which increases from bottom to top (the values are given in the caption of Fig.\ \ref{fig:profiles}). The configuration with lowest $H_{c1}$ goes from $n=10$ (bottom) successively to $n=1$ (top). 
} \label{fig:winding}
\end{center}
\end{figure}

We start by discussing individual flux tube profiles and the associated magnetic fields. We do so by choosing a fixed $T_c/\mu_q$ such that for vanishing strange quark mass there is a magnetic field at which it is energetically favorable to put a domain wall in the system. The values of $T_c/\mu_q$ for which this is the case are known from Ref.\ \cite{Haber:2017oqb} (see Fig. 5 in that reference). The domain wall interpolates between the 2SC$_{\rm ud}$ and 2SC$_{\rm us}$ 
phases, i.e., on one side, and far away from it, we have $f_3=1$, $f_2=0$, and on the other side $f_2=1$, $f_3=0$. At nonzero $m_s$ the free energies of 2SC$_{\rm ud}$ and 2SC$_{\rm us}$ are obviously no longer equal, as we have seen explicitly in Sec.\ \ref{sec:hom}, and thus the domain wall configuration becomes unstable. In Fig.\ \ref{fig:profiles} we have chosen 10 different nonzero values of $m_s/\mu_q$, such that the configuration with lowest $H_{c1}$ (to which we will refer as the "energetically preferred" or simply the "preferred" configuration)  has winding number $n=10, \ldots, 1$ as $m_s/\mu_q$ is increased.  The profiles and magnetic fields are plotted at the corresponding minimal $H_{c1}$. 
The critical fields as a function of the winding number for the parameters of Fig.\ \ref{fig:profiles} are shown in Fig.\ \ref{fig:winding}, which proves the successive decrease in the preferred winding from low to high strange quark mass.

Let us first discuss the profiles themselves in Fig.\ \ref{fig:profiles}. For the smallest masses shown here the profiles of the condensates are reminiscent of a domain wall profile: The second condensate, which is induced in the core, assumes essentially the value of the homogeneous 2SC condensate. Of course, in contrast to a domain wall, the flux tubes have a finite radius, which decreases as the winding number decreases (with increasing strange quark mass). For a quantitative discussion of the size of the flux tubes we use the various characteristic length scales obtained from the asymptotic 
behavior of the flux tube profile, see Appendix \ref{appA}. The penetration depth $\ell$ usually is characteristic for the decay of the magnetic field away from the center of the flux tube. In our case this scale corresponds to the decay of the $\tilde{B}_8$ field (in Fig.\ \ref{fig:profiles} we have restricted ourselves to plotting the ordinary magnetic field $B$). The coherence length  $\xi_3$ introduced in Eq.\ (\ref{R}) indicates the scale on which the $ud$ condensate $f_3$ recovers its homogeneous value and corresponds to the usual coherence length of a single-component flux tube. Finally, and in contrast to the standard single-component scenario, our calculation in Appendix \ref{appA} shows that there is a third scale $\xi_2$, characteristic for the induced $us$ condensate $f_2$. Using Eqs.\ (\ref{lpen}), (\ref{lcoh}),  and (\ref{lind}) we can write these three length scales in physical units as 
\begin{subequations}\allowdisplaybreaks
\label{ellxi}
\bea
\ell &\simeq& 0.94 \left[1+\frac{1}{48}\left(\frac{m_s}{\mu_q}\right)^2\ln\frac{\mu_q}{T_c}\right] \left(\frac{\mu_q}{400\, {\rm MeV}}\right)^{-1}\, {\rm fm} \, , \label{ell1} \\[2ex]
\xi_3 &\simeq& \frac{0.066}{T_c/\mu_q}   \left[1+\frac{1}{48}\left(\frac{m_s}{\mu_q}\right)^2\ln\frac{\mu_q}{T_c}\right]\left(\frac{\mu_q}{400\, {\rm MeV}}\right)^{-1}\, {\rm fm} \, , \label{xi31}  \\[2ex]
\label{lind0}
\xi_2 &\simeq& 1.3 \left(\frac{H}{\mu^2/\sqrt{\lambda}}\frac{T_c}{\mu_q}\right)^{-1/2}\left(\frac{\mu_q}{400\, {\rm MeV}}\right)^{-1} {\rm fm} \, , 
\eea
\end{subequations}
where we have used Eq.\ (\ref{weak}) with $T=0$, the 2SC condensate (\ref{rho22SC}), and $g=3.5$. The length scale $\xi_2$ on which the induced condensate varies depends on the external magnetic field $H$, for which the dimensionless value (in units of $\mu^2/\sqrt{\lambda}$) can be inserted into Eq.\ (\ref{lind0}).  
For the parameters used in Fig.\ \ref{fig:profiles} the mass terms in Eqs.\ (\ref{ell1}) and (\ref{xi31}) are negligibly small and setting  $\mu_q=400\, {\rm MeV}$ we find $\ell\simeq 0.94\, {\rm fm}$, $\xi_3 \simeq 0.77 \, {\rm fm}$, and $\xi_2\simeq (1.4-1.5)\, {\rm fm}$, while the maximal value of the radial coordinate in this figure $R=10$ corresponds to $r\simeq 7.7\,{\rm fm}$. These values, characterizing the radial extent of the core of the flux tube do not depend on the winding number $n$. In the derivation in Appendix \ref{appA} we have worked with a fixed and finite $n$ while sending the radial coordinate to infinity. However, from the full numerical profiles we see that the core of the flux tube clearly grows with increasing $n$. For instance, taking the point where the two condensates have the same value as a rough measure for the size of the core, we see that for the largest winding shown here, $n=10$, corresponding to the smallest mass, $m_s=0.046\mu_q$,  this size is $r\simeq 4.6\, {\rm fm}$. This is about six times larger than the coherence length $\xi_3$, but it also shows that it only takes a relatively small strange quark mass to reduce the size of the flux tubes from infinity (domain wall) to a few fm.

The plots in the upper panels also show the profile of the function $\sqrt{f_2^2+f_3^2}$, which is the radial coordinate of the space spanned by $(f_2,f_3)$, i.e., a domain wall can be understood as a rotation in this space from the horizontal to the vertical axis. This function illustrates the depletion of the "combined" condensate in a cylindrical layer at nonzero radius for large winding numbers and in the center of the tube for low winding numbers. This depletion reflects the structure of the magnetic field, which is shown in the lower panels. We recall that the 2SC phase (just like the CFL phase) admits a fraction of the external field $H$ even in the homogeneous phase, and this fraction is close to one for $g\gg e$. Therefore, the energetic benefit of a 2SC flux tube is to admit additional magnetic flux on top of the flux already present. As a consequence of our choice $g=3.5$, the ratio $B/H$ is  larger than 99\% already in the homogeneous 2SC phase, as one can see for instance from Eq.\ (\ref{BoverH}). The magnetic field profiles in the figure, showing $B/H$ between its minimal homogeneous 2SC value (white) and the maximal value (black) turn out to be in the range $0.9975 \lesssim B/H \lesssim 0.9995$. Therefore, the magnetic field variations are unlikely to have any physical impact in the strong coupling regime. One might also wonder about the numerical accuracy needed to resolve such tiny variations of $B/H$. However, from Eq.\ (\ref{BoverH}) it is obvious that it is the small mixing angle $\vartheta_1$ ($\sin\vartheta_1\simeq 0.05$ for $g=3.5$)  that maps the variations into a tiny interval and thus the required numerical accuracy is not as high as one might think.

The magnetic profiles show interesting features. We see that the flux tubes with higher winding (small $m_s$) have their excess magnetic field concentrated in ring-like structures. This effect has been observed in the literature before with the help of a two-component abelian Higgs model \cite{Chernodub:2010sg} and a one-component model with non-standard coupling between the condensate and the abelian gauge field \cite{Bazeia:2018hyv} (see also Refs.\ \cite{Bazeia:2018eta,Bazeia:2018fhg} for similar structures of magnetic monopoles in a non-abelian model). In these studies, the so-called Bogomolny limit was considered for simplicity, which corresponds to the transition point between type-I and type-II behavior. Here we observe the effect in a numerical evaluation of the general Ginzburg-Landau equations, and in particular we can point out regions in the phase diagram where the exotic ring-like structures are the preferred configuration (for a systematic discussion of the phase diagram see next subsection). The ring-like structure is  easy to understand: In the domain wall ($m_s=0$) the excess magnetic field is concentrated in the wall for symmetry reasons. As the strange quark mass is increased, the wall "bends" to form a flux tube with finite radius, and it is obvious for continuity reasons that for very small masses the maximum of the magnetic field is still sitting in the transition region. Then, as the sequence in Fig.\ \ref{fig:profiles} shows, the magnetic rings gradually turn into ordinary flux tubes as $m_s$ is increased (and the winding number decreases). Furthermore, we observe a double-ring structure, clearly visible in the multi-winding flux tubes. We have found that the double ring does not always occur.  For instance, for the smaller value $T_c=0.08\mu_q$ the double ring is replaced by a single ring.

\begin{figure} [t]
\begin{center}
\hbox{\includegraphics[width=0.5\textwidth]{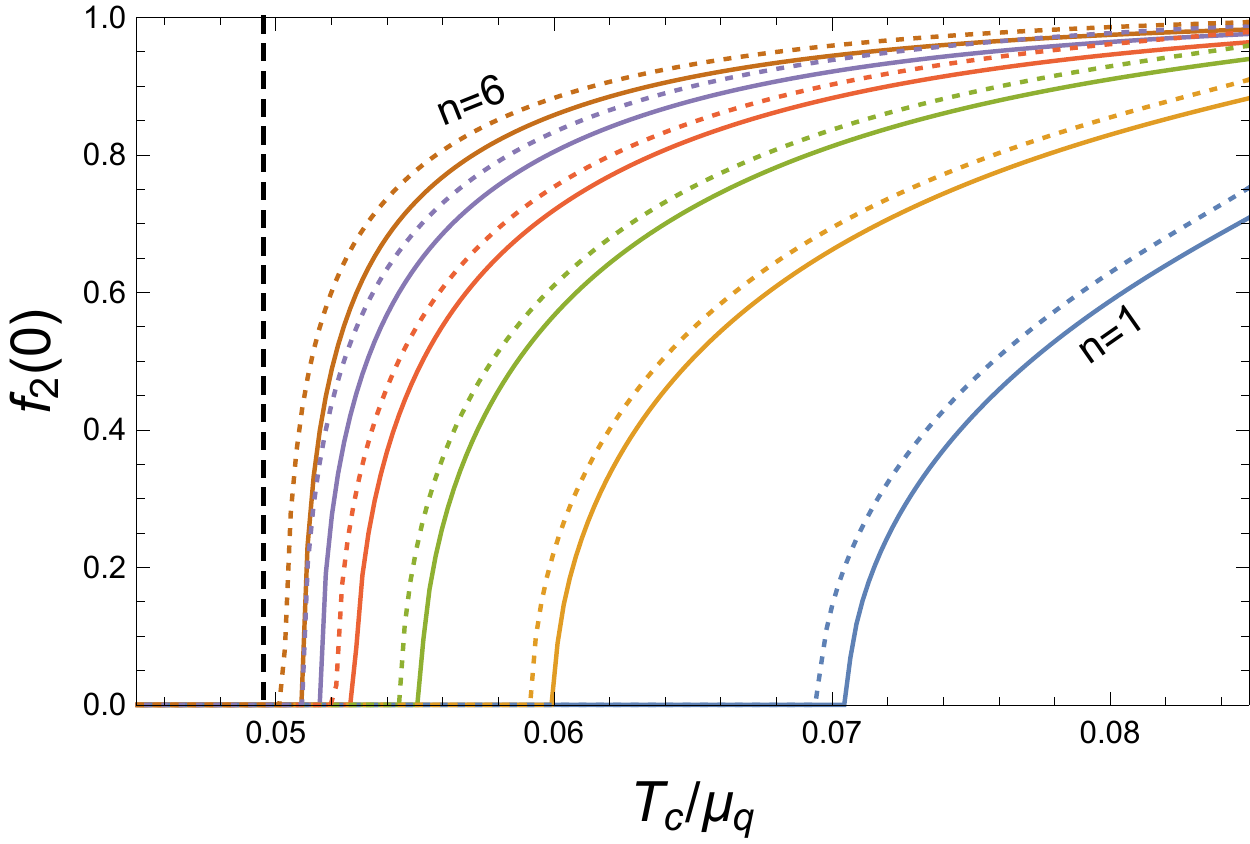}\includegraphics[width=0.5\textwidth]{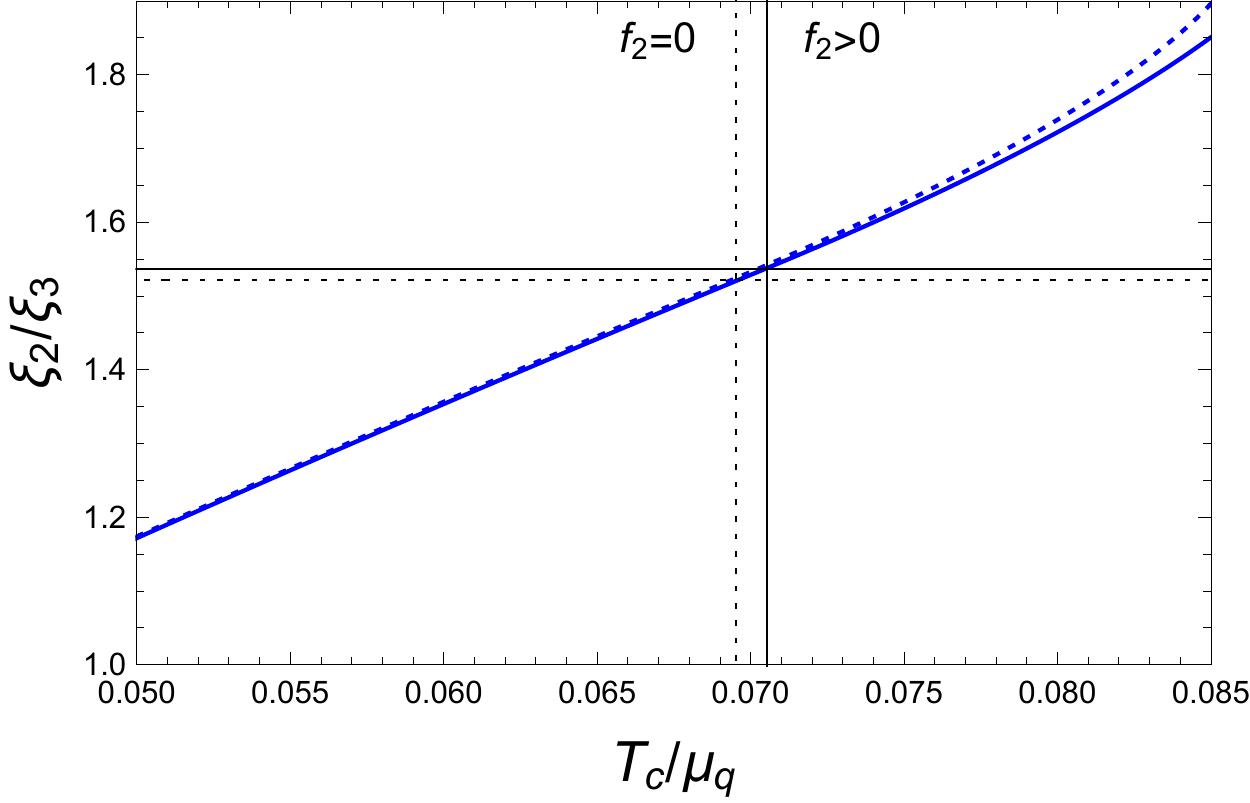}}
\caption{{\it Left panel:} Value of the dimensionless $us$ condensate $f_2(R=0)$ in the center of the 2SC flux tube as a function of $T_c/\mu_q$ at the critical field $H_{c1}$ (which depends on $T_c/\mu_q$ and $n$) for strange quark mass $m_s=0.15\mu_q$ (solid lines)  and $m_s=0$ (dotted lines), each for winding numbers $n=1,\ldots, 6$, from right to left. The standard, single-component flux tube solution exists throughout the parameter space. The vertical dashed line marks the transition point between type-I and type-II behavior. {\it Right panel:} Ratio of length scales for depleted condensate $f_3$ and induced condensate $f_2$ according to Eqs.\ (\ref{ellxi}) for the $n=1$ curves of the left panel, with the blue curves corresponding to the massless (dotted) and massive (solid) cases. The vertical lines indicate the transition from zero to nonzero induced condensate, while the horizontal lines mark the corresponding values of the ratio $\xi_2/\xi_3$ at this transition.}  \label{fig:f2}
\end{center}
\end{figure}

All profiles in Fig.\ \ref{fig:profiles} are "unconventional" in the sense that they contain two condensates. As already mentioned above, these two-component solutions are not found everywhere in  the parameter space. In the left panel of Fig.\ \ref{fig:f2} we show the value of the induced condensate in the core $f_2(0)$ as a function of the parameter $T_c/\mu_q$ for winding numbers $n=1,\ldots,6$ and for a nonzero strange quark mass, compared to the massless case. We see that as $T_c/\mu_q$ decreases, the induced condensate becomes smaller until it continuously goes to zero at a point that depends on the winding number. Interestingly, as the winding number is increased, this point seems to converge to the point (\ref{type2SCNOR}) that distinguishes type-I from type-II superconductivity. In particular, the behavior of $f_2(0)$ becomes more and more
step-like for larger windings, i.e., $f_2(0)$ is close to one until it sharply decreases near the type-I/type-II transition point. However, flux tubes with higher winding are energetically disfavored in the vicinity of the 
type-I/type-II transition point (as we shall see below), such that this interesting behavior does not seem to be physically relevant. 

The right panel of Fig.\ \ref{fig:f2} is useful to understand why the induced condensate only appears in a certain parameter regime. In this plot we show the ratio of $\xi_2$ and $\xi_3$ given in Eqs.\ (\ref{ellxi}). We see that the induced condensate vanishes if this ratio
becomes too small. For a possible interpretation note that the coherence length 
is connected to the size of the Cooper pair. If we view $\xi_3$ as a typical size of the Cooper pair (also for $us$ pairing), then the right panel of Fig.\ \ref{fig:f2} suggests that the length scale for the variation of the induced condensate, which is determined by the external magnetic field, must be sufficiently large (at least by a factor of about 1.5) compared to the size of a Cooper pair. Otherwise the condensate "does not fit" into the core of the flux tube.

%%%%%%%%%%%%%%%%%%%%%%%%%%%%%%%%%%%%%%%%%%%%%%%%%%%%%%%%%%%%%%%
\subsection{Phase structure}
\label{sec:phase}
%%%%%%%%%%%%%%%%%%%%%%%%%%%%%%%%%%%%%%%%%%%%%%%%%%%%%%%%%%%%%%%

We have seen that there are parameter choices for $(g, T_c/\mu_q, m_s/\mu_q)$ where multi-winding 2SC flux tubes with a ring-like structure of the magnetic field are energetically preferred. In this section we 
investigate the parameter space more systematically. We should keep in mind that in QCD the parameters, $g$, $T_c$, and $m_s$ are uniquely given by $\mu_q$ (at $T=0$). Therefore, as we vary the quark chemical potential, the system will move along a unique, but unknown, curve in this three-dimensional parameter space. For instance, at asymptotically large $\mu_q$, we start from small $g\ll 1$, exponentially suppressed $T_c$ and negligibly small $m_s$ (compared to $\mu_q$). Since we do not know the values of $m_s$ and $T_c$ at more moderate densities we keep our parameters as general as possible. In this sense, keeping $g=3.5$ fixed for our results is a simplification, in an even more general calculation one might also vary $g$.

\begin{figure} [t]
\begin{center}
\hbox{\includegraphics[width=0.5\textwidth]{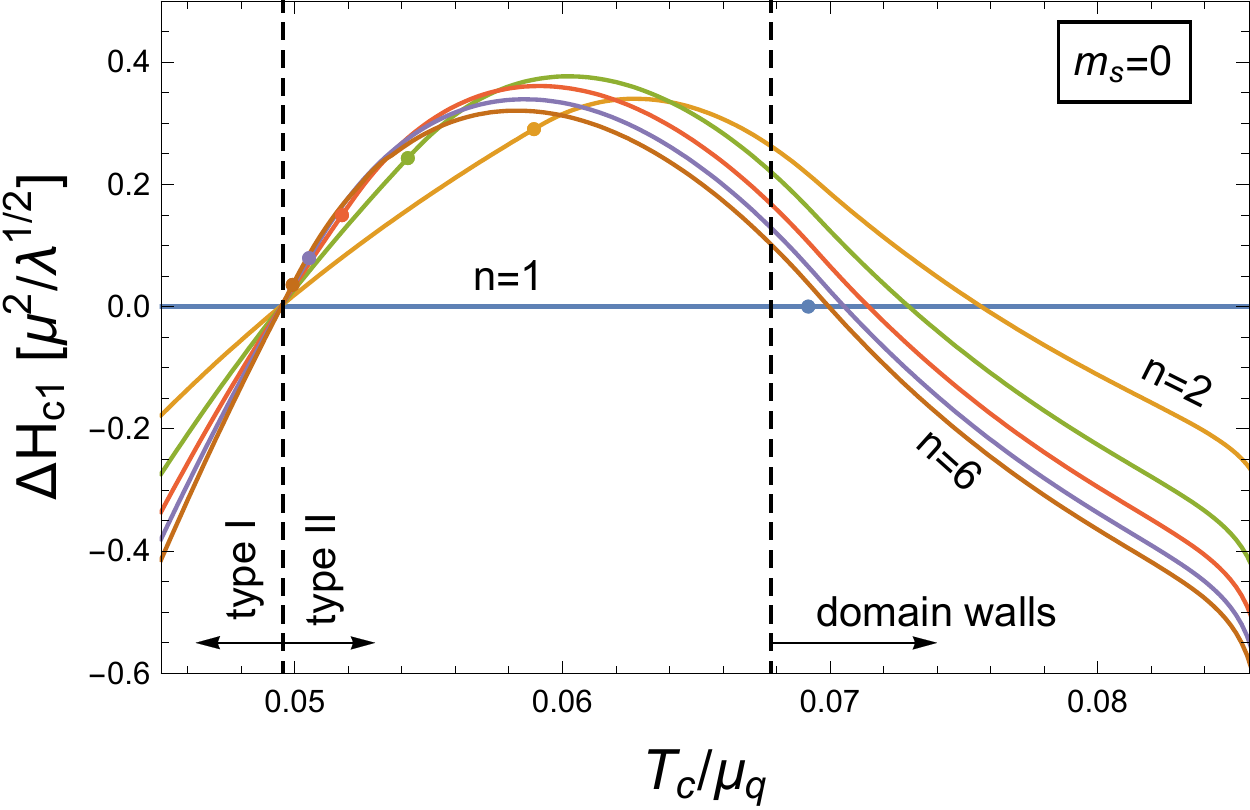}\includegraphics[width=0.5\textwidth]{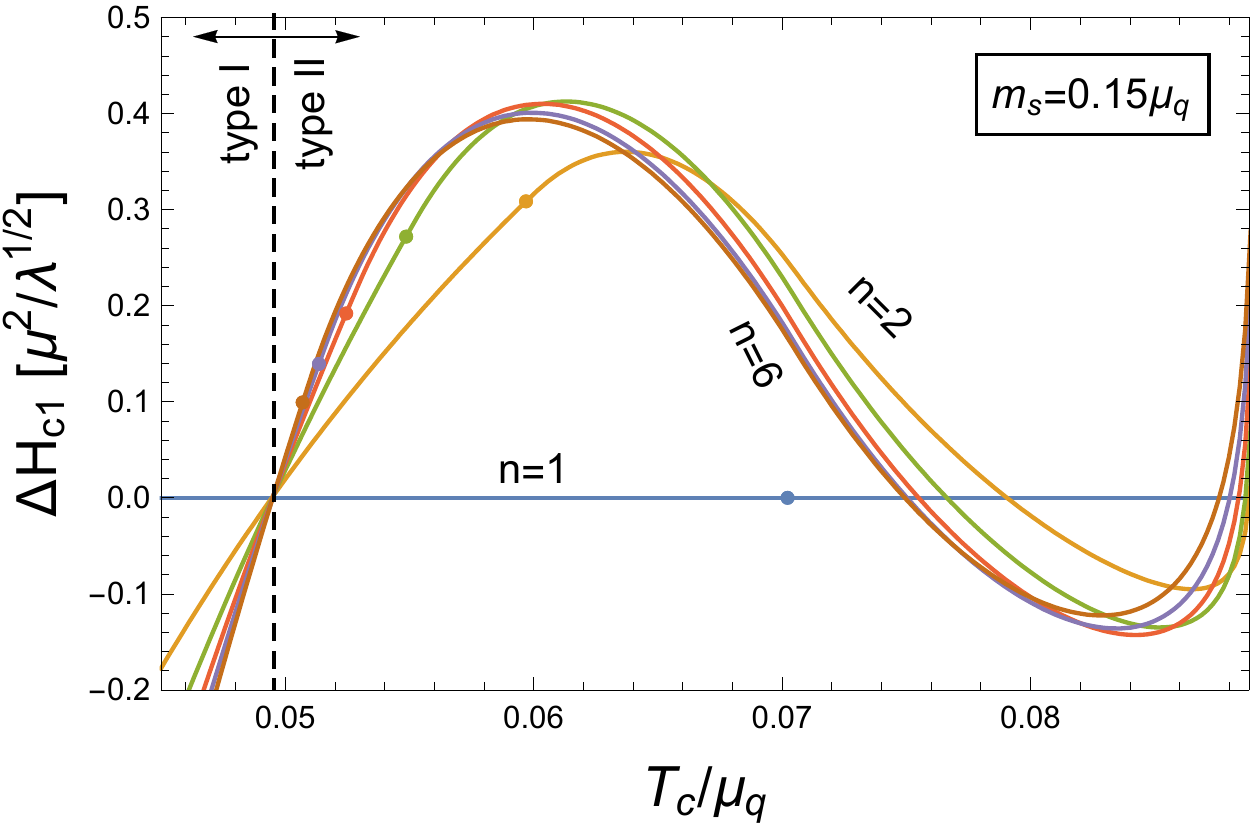}}
\caption{Critical fields for the emergence of flux tubes with winding $n$ relative to the critical field of the $n=1$ flux tube, $\Delta H_{c1} = H_{c1}(n)-H_{c1}(1)$, as a function of $T_c/\mu_q$ for the massless case (left panel) and $m_s=0.15\mu_q$ (right panel). In the massless case, domain walls ($n=\infty$) are the preferred configuration for $T_c/\mu_q \gtrsim 0.068$, as indicated by the vertical dashed line. The other vertical dashed line marks the change from type-I to type-II behavior. The dots on the curves indicate the points above which the flux tubes have a core with induced condensate $f_2$ (see also Fig.\ \ref{fig:f2}).
}\label{fig:DeltaHc1}
\end{center}
\end{figure}

In Fig.\ \ref{fig:DeltaHc1} we compare the critical magnetic fields of flux tubes with different winding numbers for two different values of $m_s$ as a function of $T_c/\mu_q$. We plot the difference of the critical fields to the critical field of the $n=1$ configuration because the different curves would be barely distinguishable had we plotted the critical fields themselves. It is instructive to start with the massless case (left panel). For $T_c/\mu_q\simeq 0.05$ we observe the standard behavior at the type-I/type-II transition: In the type-I regime, the critical field $H_{c1}$ can be lowered by increasing the winding number, and as $n\to\infty$ one expects $H_{c1}(n)$ to converge to $H_c$, the critical field at which the NOR and 2SC phases coexist. Just above the critical value $T_c/\mu_q\simeq 0.05$ the order of the different $H_{c1}(n)$ is exactly reversed, and $H_{c1}$ is given by the flux tube with lowest winding, $n=1$, the higher-winding flux tubes are disfavored.
It is a good check for our numerics that all curves intersect at the 
same point, and this point is given by Eq.\ (\ref{type2SCNOR}), whose derivation is completely independent of the numerical evaluation. This conventional behavior in the vicinity of the type-I/type-II transition point is expected because the second condensate plays no role here, at least for the lowest winding numbers, as we have seen in Fig.\ \ref{fig:f2}.

Now, as $T_c/\mu_q$ is increased, the unconventional 2SC behavior becomes apparent. For any winding $n>1$ there is a critical $T_c/\mu_q$ at which this higher winding becomes favored over $n=1$. Similar to the type-I regime, there is a region in which $H_{c1}(n)$ decreases monotonically with $n$ until, for $n\to \infty$,  it is expected to intersect $H_{c1}(1)$ at the critical point where domain walls set in (we have only plotted the curves for a few winding numbers, obviously the numerics become more challenging for large $n$). This point, $T_c/\mu_q \simeq 0.068$, is taken from  Ref.\ \cite{Haber:2017oqb}, where it was computed explicitly by using a planar instead of a cylindrical geometry  (see Figs.\ 4 and 5 in that reference). Here we have marked the onset of domain walls by a vertical dashed line.  

In the right panel we first note that the behavior in the vicinity of the type-I/type-II transition in the presence of a strange quark mass is qualitatively the same as in the massless case. The transition point itself, as we already know from Eq.\ (\ref{type2SCNOR}), is even exactly the same, at least up to lowest nontrivial order  in $m_s$. Now, however, the behavior for larger $T_c/\mu_q$, where the second condensate does play a role, is qualitatively different. For the chosen value of $m_s/\mu_q$, we find that for windings $n\le 6$ there is a regime in which flux tubes with winding $n$ are preferred, while flux tubes with $n > 6$ are never preferred.  We see that the lowest $H_{c1}$ corresponds to flux tubes with winding numbers 1,6,5,4,3,2, as $T_c/\mu_q$ is increased. These structures can be viewed as "remnants" of the domain wall.

\begin{figure} [t]
\begin{center}
\hbox{\includegraphics[width=0.5\textwidth]{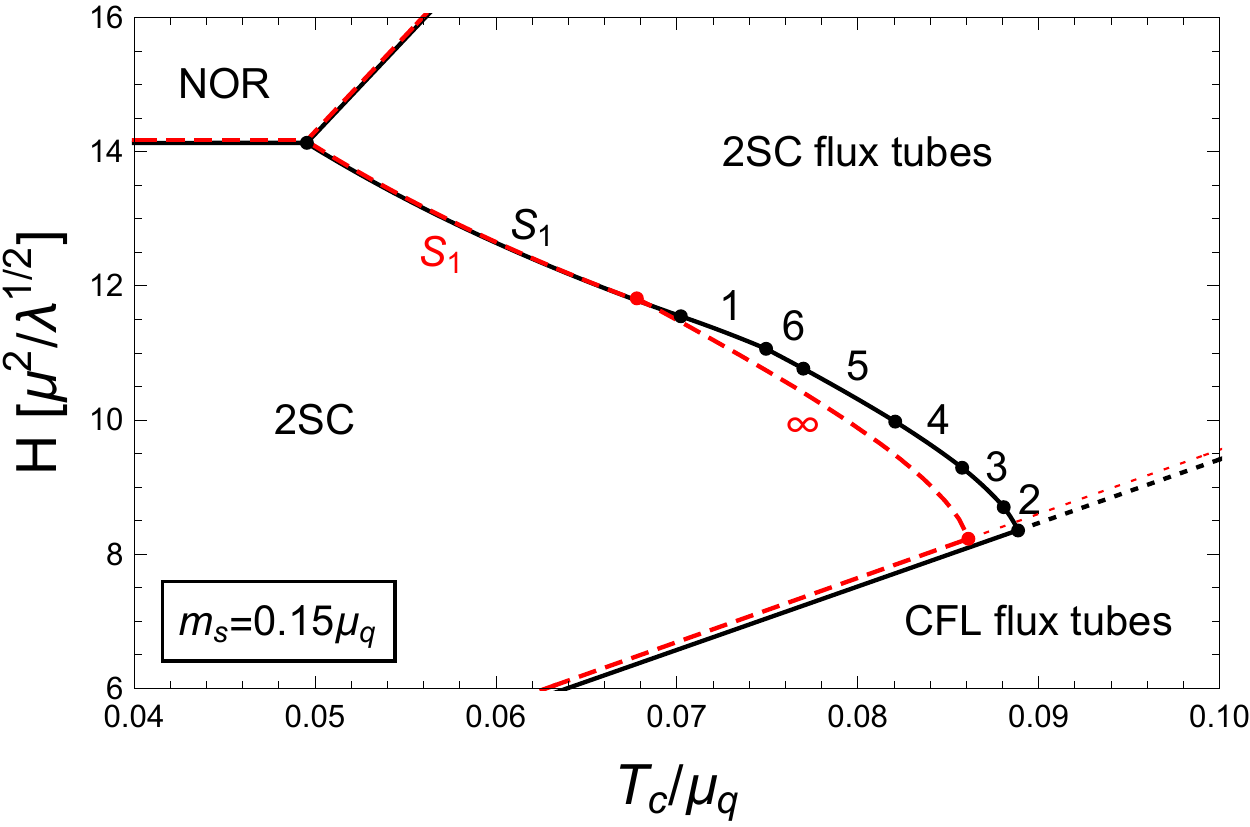}
\includegraphics[width=0.5\textwidth]{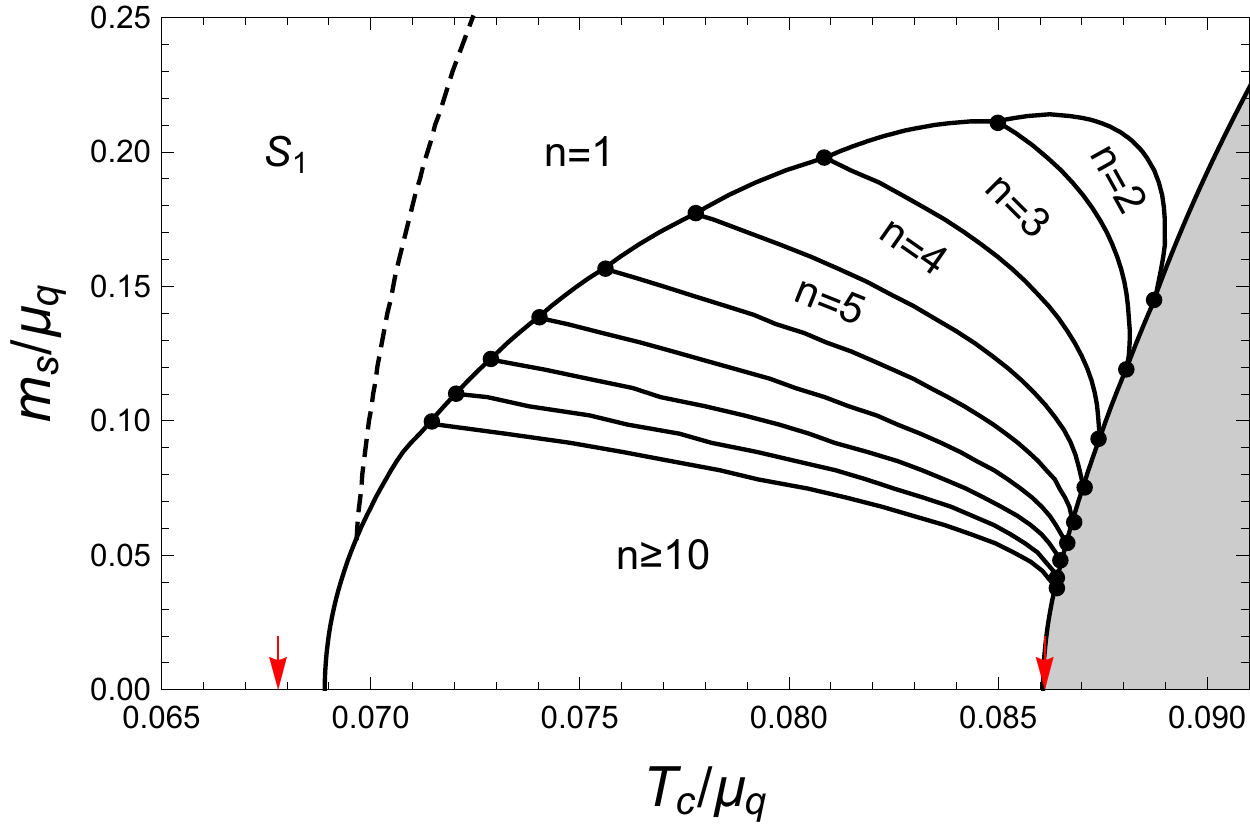}}
\caption{{\it Left panel:} Phase diagram in the plane of external field $H$ and $T_c/\mu_q$ for $m_s=0.15\mu_q$ (black solid lines) compared to the massless result from Ref.\ \cite{Haber:2017oqb} (red dashed lines). The numbers at the critical field for the emergence of 2SC flux tubes with $us$ core indicate the winding number $n$, with $n=\infty$ indicating the domain wall in the massless case. The critical lines for the emergence of 2SC flux tubes without a $us$ core (which always have winding $n=1$) are marked with $S_1$.
{\it Right panel:} Phase diagram in the $m_s/\mu_q$-$T_c/\mu_q$ plane indicating the preferred flux tube configurations. In the shaded region dSC and CFL flux tubes appear, and the red arrows indicate the range where, at $m_s=0$, domain walls are preferred. 
This phase diagram shows one of the main results of the paper, namely that remnants of the domain wall in the form of multi-winding flux tubes survive up to $m_s\simeq 0.21\mu_q$.  
}\label{fig:phasediagram}
\end{center}
\end{figure}

With the help of the results of this panel we can construct the phase diagram in the $H$-$T_c/\mu_q$ plane for this particular mass $m_s=0.15\mu_q$. To this end, we need to bring together the critical fields $H_{c1}$ from Fig.\ \ref{fig:DeltaHc1} with the critical fields $H_c$ from Sec.\ \ref{sec:Hc} and the critical fields $H_{c2}$ from Sec.\ \ref{sec:Hc2}. The result is shown in the left panel of Fig.\ \ref{fig:phasediagram}. This panel contains the first-order transition between NOR and 2SC phases (\ref{Hcs}), the second-order transitions from the NOR to the "2SC flux tube" phase  (\ref{Hc22SCNOR}) and from the 2SC to the "CFL flux tube" phase (\ref{Hc2one}), and the second-order transition from the 2SC to the "2SC flux tube" phase just discussed, including the different segments corresponding to different winding numbers. We have included the phase transitions of the massless case for comparison, including the segment where domain walls are formed. The single-component flux tube configuration, i.e., the one without a $us$ condensate in the core, is denoted by $S_1$, following the notation of Ref.\ \cite{Haber:2017oqb}. This configuration always has winding number 1.  The region labeled by "CFL flux tubes" is bounded from above by $H_{c2}^{(1)}$, which actually indicates a transition to dSC flux tubes (see discussion below Eqs.\ (\ref{Hc212})). For the given parameters, the "would-be" critical magnetic field $H_{c2}^{(2)}$ is very close to $H_{c2}^{(1)}$, and thus we have, slightly inaccurately, termed the entire region "CFL flux tubes", although we know that there is at least a thin slice where the inhomogeneous phase is not made of CFL flux tubes. 

Since our calculation only allows us to determine the critical fields for the entrance into the flux tube phases, we can only speculate about the structure of these inhomogeneous phases away from the second-order lines. It is obvious that the structure will be more complicated than in a standard single-component superconductor. This, firstly, concerns the transition between the phases labeled by "CFL flux tubes"  and "2SC flux tubes". We have continued the $H_{c2}$ transition curve into the flux tube regime as a thin dotted line, but of course this curve should not be taken too seriously because it was calculated under the assumption of a homogeneous 2SC phase on one side of the transition. Secondly, and more closely related to our main results, the different winding numbers that occur upon entering the 2SC flux tube phase also suggest a complicated structure of the flux tube arrays. It is conceivable that there are transitions between "pure" arrays of a single winding number or that there are arrays composed of flux tubes with different winding numbers.  

We may now ask up to which values of the strange quark mass the multi-winding flux tubes survive. More precisely, at which value of $m_s$ do only $n=1$ flux tubes form at the entire phase transition curve $H_{c1}$ between the homogeneous 2SC phase and the "2SC flux tube" phase? This question can be answered by repeating the calculation needed for left panel of Fig.\ \ref{fig:phasediagram} for different values of $m_s$ and determining the sequence of the preferred configurations as a  function of $T_c/\mu_q$ in each case. The result is shown in in the right panel of Fig.\ \ref{fig:phasediagram}. In this panel, the $m_s/\mu_q$-$T_c/\mu_q$ plane is divided into different regions, each region labeled by the preferred flux tube configuration. One can check that a slice through this plot at $m_s=0.15\mu_q$ reproduces exactly the sequence of phases shown in the left panel. In particular, we have indicated the transition between the standard flux tubes $S_1$ and the two-component flux tubes with winding number $n=1$ by a dashed curve. We have restricted our calculation to winding numbers $n\le 10$, but it is obvious how the pattern continues: As we move towards $m_s=0$, say at fixed $T_c/\mu_q$, the winding number of the preferred configuration increases successively and more rapidly. Eventually, the winding number and thickness of the flux tube go to infinity,  which corresponds to the domain wall, whose range is indicated with (red) arrows. Therefore, the lines that bound the region of multi-winding configurations
at small $m_s$ from both sides -- here calculated from $n=10$ -- will slightly change if higher windings are taken into account and they are expected to converge to the arrows at $m_s=0$. In the shaded region, dSC and CFL flux tubes start to become relevant, its boundary is given by the condition $H_{c1}=H_{c2}^{(1)}$ (see also the discussion around Eq.\ (\ref{Xic2})).

The main conclusion of this phase diagram is that the maximal strange quark mass for which remnants of the domain wall (i.e., 2SC flux tubes with winding number larger than one) exist, is $m_s\simeq 0.21\mu_q$. This is on the lower end of the range expected for compact stars. Therefore, at least within our approximations, we conclude that it is conceivable, but unlikely that 2SC flux tubes with exotic structures as shown in Fig.\ \ref{fig:profiles} play an important role in the astrophysical setting. However, our phase diagram shows that two-component flux tubes (with $n=1$) do survive for larger values of $m_s$. Since they have a $us$ condensate in their core, they can also be considered as remnants of the domain wall, albeit with  conventional magnetic field structure. We have checked that if the phase diagram in the right panel of Fig.\ \ref{fig:phasediagram} is continued to about $m_s=0.5\mu_q$, there is still a sizable range (of roughly the same size as for $m_s=0.25\mu_q$), starting at $T_c/\mu_q\simeq 0.083$, where 2SC flux tubes with $us$ core are preferred. Consequently, they may well have to be taken into account for the physics of compact stars.

In the discussion of the applicability of our results to compact stars, we should also return to the actual magnitude of the magnetic fields considered here. Using the conversion to physical units given below Eq.\ (\ref{triple2}) and the expressions for $\mu$ and $\lambda$ in Eq.\ (\ref{weak}), we find that the magnetic field in Gauss can be computed from 
\be
H = 1.6\times 10^{19} \frac{H}{\mu^2/\sqrt{\lambda}}\ \frac{T_c}{\mu_q} \left(1-\frac{T}{T_c}\right)\left(\frac{\mu_q}{400\, {\rm MeV}}\right)^2 G \, .
\ee
Consequently, setting $T=0$, the triple point at $T_c/\mu_q \simeq 0.05$ in the left panel of Fig.\ \ref{fig:phasediagram}, where NOR, 2SC, and 2SC flux tube phases meet, corresponds to about $H\simeq 1.1\times 10^{19}\, {\rm G}$, while the critical fields at $T_c/\mu_q=0.08$ are $H_{c1}\simeq 1.3 \times 10^{19}\, {\rm G}$ and $H_{c2}\simeq 2.9\times 10^{19}\, {\rm G}$ (note that $H$ 
in physical units also varies along the {\it horizontal} direction of this plot since $\mu^2/\lambda^{1/2}$ depends on $T_c/\mu_q$). Magnetic fields of this magnitude are getting close to the limit for the stability of the star, and they are several orders of magnitude larger than observed at the surface. It is thus highly speculative, although possible, that these huge magnetic fields are reached in the interior of the star. However, exotic flux tubes as discussed here may be formed dynamically, for instance if the star cools through the superconducting phase transition at a roughly constant magnetic field \cite{Alford:2010qf}. And, if  color-magnetic flux tubes exist in compact stars, they  help to sustain possible ellipticities of the star, resulting in detectable gravitational waves \cite{Glampedakis:2012qp}. 
Therefore, even though the equilibrium situation studied here may never be reached in an astrophysical setting, it is important to understand the various possible -- non-standard -- flux tube configurations. 

%%%%%%%%%%%%%%%%%%%%%%%%%%%%%%%%%%%%%%%%%%%%%%%%%%%%%%%%%%%%%%%
\section{Summary and outlook}
\label{sec:summary}
%%%%%%%%%%%%%%%%%%%%%%%%%%%%%%%%%%%%%%%%%%%%%%%%%%%%%%%%%%%%%%%

Using a Ginzburg-Landau approach, we have investigated magnetic flux tubes in the 2SC phase of dense quark matter, i.e., flux tubes that asymptote far away from their center to a phase where $u$ and $d$ quarks form a Cooper pair condensate. 
Improving earlier studies, we have included the strange quark mass $m_s$ as a free parameter in the microscopic calculation of the flux tube profiles and their free energies. This is more realistic and makes a qualitative difference: In the presence of the strange quark mass, domain wall configurations that interpolate between $ud$ and $us$ condensates
and that have previously been found to be energetically favored in a certain parameter region are no longer stable. We have shown that these domain walls -- which can be viewed as flux tubes with infinite radius -- turn into flux tubes with finite radius and high winding number as the strange quark mass is switched on. These unconventional flux tubes, which have a $us$ condensate in their core, exhibit a ring-like structure of the magnetic field, i.e., the maximum of the magnetic field sits at a nonzero value of the radial distance from the center. Already for moderate values $m_s\simeq 0.05\mu_q$, the flux tube radius is reduced to under $5\, {\rm fm}$, and flux tubes with winding number larger than one survive up to about $m_s\simeq 0.2\mu_q$. This roughly corresponds to the lower limit that is expected for the effective strange quark mass  in a compact star environment. Therefore, at least within our approximations, we have concluded that the exotic multi-winding flux tubes -- "remnants" of the domain walls from the massless limit  -- probably play no major role in compact stars, although flux tubes with a $us$ core, but winding number one, survive up to much larger masses. Also, since we have only addressed equilibrium configurations, all our magnetic defects require magnetic field strengths of the order of $10^{19}\, {\rm G}$, which are unlikely to be reached in compact stars, while it is conceivable that the defects exist as a metastable configuration during the evolution of the star. Our study is also relevant for the QCD phase diagram at a nonzero external magnetic field and for a more general understanding of two-component magnetic flux tubes. Although the situation in dense quark matter is very specific due to the presence of color-magnetic fields and their mixing with the ordinary magnetic field, our 2SC flux tubes with ring-like magnetic fields are not unlike configurations previously found in different models  \cite{Chernodub:2010sg,Bazeia:2018hyv}. 

Our results suggest  a very rich structure of the 2SC flux tube phase for not too large values of $m_s$, containing flux tubes of different winding numbers. It would be interesting to see if there are phase transitions between different flux tube arrays or if there are arrays composed of flux tubes with different winding. We have also pointed out that the upper critical field of the CFL flux tubes, i.e., the transition to a homogeneous 2SC phase, is affected qualitatively by the strange quark mass. Just below this upper critical field dSC flux tubes seem to be favored, such that also the CFL flux tubes phase appears to be more complicated than expected from the massless limit. Also the CFL flux tubes themselves may be revisited using our setup in order to systematically investigate the effect of the strange quark mass. In this case, there is no equivalent of the domain wall and thus the effect is probably less dramatic than in the 2SC phase.

We have used various approximations that can be improved  in future studies. It would be interesting, but also very challenging, to start from a fermionic approach rather than from the bosonic Ginzburg-Landau model. This may modify our results since the magnetic fields considered here are sufficiently large to resolve the fermionic structure of the Cooper pairs. A more straightforward extension would be to explore the parameter space more systematically. We have restricted ourselves to the weak-coupling values for the parameters of the Ginzburg-Landau potential, and extrapolated the results to a large strong coupling constant realistic for compact stars. Also, we have neglected off-diagonal components in the order parameter matrix, which become potentially relevant due to the explicitly broken flavor symmetry. It would also be interesting to improve the  treatment of temperature in the Ginzburg-Landau potential used here and in  previous studies. Since there are several condensates which are not expected to melt at the same point, a more refined temperature dependence would include more than one critical temperature \cite{Haber:2017kth}.

\begin{acknowledgments}
We thank A.\ Haber for valuable comments and discussions. A.S.\ is supported by the Science \& Technology Facilities Council (STFC) in the form of an Ernest Rutherford Fellowship.
\end{acknowledgments}

\appendix

%%%%%%%%%%%%%%%%%%%%%%%%%%%%%%%%%%%%%%%%%%%%%%%%%%%%%%%%%%%%%%%%%%%%%%%%%%%%%%%%%%%
\section{Asymptotics for large distances}
\label{appA}
%%%%%%%%%%%%%%%%%%%%%%%%%%%%%%%%%%%%%%%%%%%%%%%%%%%%%%%%%%%%%%%%%%%%%%%%%%%%%%%%%%

In this appendix, we discuss the asymptotic behavior of the condensates and the 
gauge fields far away from the center of the flux tube, in particular the non-standard behavior of the
 condensate induced in the core, i.e., the condensate that vanishes at infinite radial distance. We restrict ourselves to the situation of two non-vanishing condensates, the $ud$ condensate $f_3$, which assumes its homogeneous value at large distances, and the induced $us$ condensate $f_2$. The third condensate, $f_1$, is omitted because it never plays a role in the preferred 2SC flux tube configurations. As in the analogous calculation of CFL flux tubes \cite{Haber:2017oqb,Haber:2018tqw} we use the 
following ansatz for the gauge fields and the condensates,
\begin{subequations} \label{asymp}
\bea 
\tilde{a}_3(R) &=& \tilde{a}_3(\infty)+R\tilde{v}_3(R) \, , \qquad  \tilde{a}_8(R) = \frac{n}{\tilde{q}_{83}}+R\tilde{v}_8(R) \, , \\[2ex]
f_2(R) &=& u_2(R) \, , \qquad f_3(R) = 1+u_3(R) \, ,
\eea
\end{subequations}
where we have taken into account the boundary conditions at $R=\infty$ for $\tilde{a}_8$, $f_2$, and $f_3$, while the boundary value for $\tilde{a}_3$ is a dynamical quantity. We have introduced the functions $\tilde{v}_3$, $\tilde{v}_8$, $u_2$, $u_3$, which all approach 0 as $R\to\infty$. We assume (and this will be confirmed a posteriori) that they fall off exponentially at large $R$. 

With this ansatz we find that the equation of motion for $f_2$ (\ref{eqf2})
becomes
\be \label{u2approx}
0\simeq u_2''+\frac{u_2'}{R}+u_2(-\Xi^2R^2+\ldots) \, ,
\ee
where the dots stand for terms that are suppressed exponentially or by powers of $R$ compared to the $\Xi^2R^2$ term. This term thus plays a crucial role in the asymptotic behavior,  and we find the asymptotic solution 
\be \label{u2asymp}
u_2(R) \simeq \frac{d_2}{R}e^{-\Xi R^2/2} \, , 
\ee
where $d_2$ is an integration constant which can only be computed from solving the 
full differential equations numerically. The solution (\ref{u2asymp}) fulfills 
Eq.\ (\ref{u2approx}) if we neglect terms that are suppressed by powers of $R$
compared to the leading-order terms.  Note that here and in the rest of this appendix we treat the winding number $n$ as fixed and finite. A different analysis, carefully taking into account the two limits $R\to\infty$ and $n\to \infty$ would be needed for the transition from flux tube to domain wall solutions, see Ref.\ \cite{Penin:2020cxj} for a recent study in the abelian Higgs model. 

Next, we consider the equation of motion for $\tilde{a}_3$ (\ref{a3t}), which becomes
\be
R\tilde{v}_3''+\tilde{v}_3'-\frac{\tilde{v}_3}{R} \simeq \frac{\tilde{q}_3d_2^2}{\lambda R^2}(\Xi^2 R^2+\ldots )e^{-\Xi R^2}\, .
\ee
Again only keeping the leading-order contributions we find that the asymptotic behavior of the gauge field $\tilde{a}_3$ is given by
 \be
\tilde{v}_3 \simeq  \frac{\tilde{q}_3d_2^2}{4\lambda \Xi R^3}e^{-\Xi R^2} \, . 
\ee
We see that the condensate $f_2$ and the deviation of the gauge field $\tilde{a}_3$ from $\tilde{a}_3(\infty)$ are both suppressed by $\exp(-{\rm const}\times R^2)$. This is in contrast to the behavior of a single-component flux tube, where the 
suppression is $\exp(-{\rm const}\times R)$, which, as we shall now see, is assumed by the functions $\tilde{a}_8$ and $f_3$. Anticipating the milder suppression of $\tilde{v}_8$ and $u_3$, i.e., neglecting terms of order $\exp(-{\rm const}\times R^2)$, we can approximate the remaining equations of motion (\ref{a8t}) and (\ref{eqf3}) by 
\begin{subequations}
\bea
0&\simeq& R^2 \tilde{v}_8''+R\tilde{v}_8'-\tilde{v}_8\left(1+\frac{\tilde{q}_{83}^2}{\lambda}R^2\right)  \, , \\[2ex] 
0&\simeq& R^2u_3''+Ru_3'-2u_3R^2 \, .
\eea
\end{subequations}
These equations are identical to the standard single-component case, and their solution can be written in terms of modified Bessel functions of the second kind. Here we simply quote the leading asymptotic behavior,
\be
\tilde{v}_8(R) \simeq \frac{\tilde{c}_8}{\sqrt{R}} e^{-\tilde{q}_{83}R/\sqrt{\lambda}}\, , 
\qquad u_3(R) = \frac{d_3}{\sqrt{R}} e^{-\sqrt{2}R} \, ,
\ee
where $\tilde{c}_8$ and $d_3$ are constants that can be determined from the full numerical solution. Recalling the definition of the dimensionless radial 
coordinate $R$ (\ref{R}), we see that this results indicates the usual behavior: The 
length scale for the exponential decay of the rotated magnetic field $\tilde{B}_8$ is given by 
the penetration depth,
\be \label{lpen}
\frac{\tilde{q}_{83}R}{\sqrt{\lambda}} = \frac{r}{\ell} \, , \qquad \ell = \frac{1}{\tilde{q}_{83}\rho_{\rm 2SC}} \, , 
\ee
and the length scale on which the condensate approaches its homogeneous value is the coherence length, already introduced in Eq.\ (\ref{R}),
\be \label{lcoh}
R = \frac{r}{\xi_3}  \, , \qquad \xi_3= \frac{1}{\sqrt{\lambda}\rho_{\rm 2SC}} \, . 
\ee
In contrast, the induced condensate decays on a different length scale, given by the external magnetic field, 
\be \label{lind}
\Xi^{1/2}R = \frac{r}{\xi_2}  \, , \qquad \xi_2 = 
\left(\frac{\tilde{q}_3H\cos\vartheta_1\sin\vartheta_2}{2}\right)^{-1/2} = \left[\frac{3eg^2 H}{4(3g^2+e^2)}\right]^{-1/2} \, ,
\ee
where we have used the definition of the dimensionless magnetic field $\Xi$ (\ref{Xi}). We have added the subscript 2 to indicate that it is the condensate $f_2$ that varies on the length scale $\xi_2$.

\bibliography{references}

\begin{thebibliography}{61}
\expandafter\ifx\csname natexlab\endcsname\relax\def\natexlab#1{#1}\fi
\expandafter\ifx\csname bibnamefont\endcsname\relax
  \def\bibnamefont#1{#1}\fi
\expandafter\ifx\csname bibfnamefont\endcsname\relax
  \def\bibfnamefont#1{#1}\fi
\expandafter\ifx\csname citenamefont\endcsname\relax
  \def\citenamefont#1{#1}\fi
\expandafter\ifx\csname url\endcsname\relax
  \def\url#1{\texttt{#1}}\fi
\expandafter\ifx\csname urlprefix\endcsname\relax\def\urlprefix{URL }\fi
\providecommand{\bibinfo}[2]{#2}
\providecommand{\eprint}[2][]{\url{#2}}

\bibitem[{\citenamefont{Alford et~al.}(2008)\citenamefont{Alford, Schmitt,
  Rajagopal, and Sch{\"a}fer}}]{Alford:2007xm}
\bibinfo{author}{\bibfnamefont{M.~G.} \bibnamefont{Alford}},
  \bibinfo{author}{\bibfnamefont{A.}~\bibnamefont{Schmitt}},
  \bibinfo{author}{\bibfnamefont{K.}~\bibnamefont{Rajagopal}},
  \bibnamefont{and}
  \bibinfo{author}{\bibfnamefont{T.}~\bibnamefont{Sch{\"a}fer}},
  \bibinfo{journal}{Rev.Mod.Phys.} \textbf{\bibinfo{volume}{80}},
  \bibinfo{pages}{1455} (\bibinfo{year}{2008}), \eprint{0709.4635}.

\bibitem[{\citenamefont{Bali et~al.}(2012)\citenamefont{Bali, Bruckmann,
  Endrodi, Fodor, Katz, Krieg, Sch{\"a}fer, and Szabo}}]{Bali:2011qj}
\bibinfo{author}{\bibfnamefont{G.}~\bibnamefont{Bali}},
  \bibinfo{author}{\bibfnamefont{F.}~\bibnamefont{Bruckmann}},
  \bibinfo{author}{\bibfnamefont{G.}~\bibnamefont{Endrodi}},
  \bibinfo{author}{\bibfnamefont{Z.}~\bibnamefont{Fodor}},
  \bibinfo{author}{\bibfnamefont{S.}~\bibnamefont{Katz}},
  \bibinfo{author}{\bibfnamefont{S.}~\bibnamefont{Krieg}},
  \bibinfo{author}{\bibfnamefont{A.}~\bibnamefont{Sch{\"a}fer}},
  \bibnamefont{and} \bibinfo{author}{\bibfnamefont{K.}~\bibnamefont{Szabo}},
  \bibinfo{journal}{JHEP} \textbf{\bibinfo{volume}{02}}, \bibinfo{pages}{044}
  (\bibinfo{year}{2012}), \eprint{1111.4956}.

\bibitem[{\citenamefont{Kharzeev et~al.}(2013)\citenamefont{Kharzeev,
  Landsteiner, Schmitt, and Yee}}]{Kharzeev:2013jha}
\bibinfo{author}{\bibfnamefont{D.}~\bibnamefont{Kharzeev}},
  \bibinfo{author}{\bibfnamefont{K.}~\bibnamefont{Landsteiner}},
  \bibinfo{author}{\bibfnamefont{A.}~\bibnamefont{Schmitt}}, \bibnamefont{and}
  \bibinfo{author}{\bibfnamefont{H.-U.} \bibnamefont{Yee}},
  \bibinfo{journal}{Lect.Notes Phys.} \textbf{\bibinfo{volume}{871}},
  \bibinfo{pages}{1} (\bibinfo{year}{2013}).

\bibitem[{\citenamefont{Kharzeev et~al.}(2008)\citenamefont{Kharzeev, McLerran,
  and Warringa}}]{Kharzeev:2007jp}
\bibinfo{author}{\bibfnamefont{D.~E.} \bibnamefont{Kharzeev}},
  \bibinfo{author}{\bibfnamefont{L.~D.} \bibnamefont{McLerran}},
  \bibnamefont{and} \bibinfo{author}{\bibfnamefont{H.~J.}
  \bibnamefont{Warringa}}, \bibinfo{journal}{Nucl. Phys. A}
  \textbf{\bibinfo{volume}{803}}, \bibinfo{pages}{227} (\bibinfo{year}{2008}),
  \eprint{0711.0950}.

\bibitem[{\citenamefont{Adamczyk et~al.}(2015)}]{Adamczyk:2015eqo}
\bibinfo{author}{\bibfnamefont{L.}~\bibnamefont{Adamczyk}} \bibnamefont{et~al.}
  (\bibinfo{collaboration}{STAR}), \bibinfo{journal}{Phys. Rev. Lett.}
  \textbf{\bibinfo{volume}{114}}, \bibinfo{pages}{252302}
  (\bibinfo{year}{2015}), \eprint{1504.02175}.

\bibitem[{\citenamefont{{Lai} and {Shapiro}}(1991)}]{1991ApJ...383..745L}
\bibinfo{author}{\bibfnamefont{D.}~\bibnamefont{{Lai}}} \bibnamefont{and}
  \bibinfo{author}{\bibfnamefont{S.~L.} \bibnamefont{{Shapiro}}},
  \bibinfo{journal}{\apj} \textbf{\bibinfo{volume}{383}}, \bibinfo{pages}{745}
  (\bibinfo{year}{1991}).

\bibitem[{\citenamefont{Ferrer et~al.}(2010)\citenamefont{Ferrer, de~la Incera,
  Keith, Portillo, and Springsteen}}]{Ferrer:2010wz}
\bibinfo{author}{\bibfnamefont{E.~J.} \bibnamefont{Ferrer}},
  \bibinfo{author}{\bibfnamefont{V.}~\bibnamefont{de~la Incera}},
  \bibinfo{author}{\bibfnamefont{J.~P.} \bibnamefont{Keith}},
  \bibinfo{author}{\bibfnamefont{I.}~\bibnamefont{Portillo}}, \bibnamefont{and}
  \bibinfo{author}{\bibfnamefont{P.~L.} \bibnamefont{Springsteen}},
  \bibinfo{journal}{Phys. Rev. C} \textbf{\bibinfo{volume}{82}},
  \bibinfo{pages}{065802} (\bibinfo{year}{2010}), \eprint{1009.3521}.

\bibitem[{\citenamefont{Potekhin and Yakovlev}(2012)}]{Potekhin:2011eb}
\bibinfo{author}{\bibfnamefont{A.}~\bibnamefont{Potekhin}} \bibnamefont{and}
  \bibinfo{author}{\bibfnamefont{D.}~\bibnamefont{Yakovlev}},
  \bibinfo{journal}{Phys. Rev. C} \textbf{\bibinfo{volume}{85}},
  \bibinfo{pages}{039801} (\bibinfo{year}{2012}), \eprint{1109.3783}.

\bibitem[{\citenamefont{Bocquet et~al.}(1995)\citenamefont{Bocquet, Bonazzola,
  Gourgoulhon, and Novak}}]{Bocquet:1995je}
\bibinfo{author}{\bibfnamefont{M.}~\bibnamefont{Bocquet}},
  \bibinfo{author}{\bibfnamefont{S.}~\bibnamefont{Bonazzola}},
  \bibinfo{author}{\bibfnamefont{E.}~\bibnamefont{Gourgoulhon}},
  \bibnamefont{and} \bibinfo{author}{\bibfnamefont{J.}~\bibnamefont{Novak}},
  \bibinfo{journal}{Astron. Astrophys.} \textbf{\bibinfo{volume}{301}},
  \bibinfo{pages}{757} (\bibinfo{year}{1995}), \eprint{gr-qc/9503044}.

\bibitem[{\citenamefont{Schmitt}(2016)}]{Schmitt:2016pre}
\bibinfo{author}{\bibfnamefont{A.}~\bibnamefont{Schmitt}},
  \bibinfo{journal}{Eur. Phys. J. A} \textbf{\bibinfo{volume}{52}},
  \bibinfo{pages}{226} (\bibinfo{year}{2016}).

\bibitem[{\citenamefont{Bailin and Love}(1984)}]{Bailin:1983bm}
\bibinfo{author}{\bibfnamefont{D.}~\bibnamefont{Bailin}} \bibnamefont{and}
  \bibinfo{author}{\bibfnamefont{A.}~\bibnamefont{Love}},
  \bibinfo{journal}{Phys. Rept.} \textbf{\bibinfo{volume}{107}},
  \bibinfo{pages}{325} (\bibinfo{year}{1984}).

\bibitem[{\citenamefont{Alford et~al.}(1999)\citenamefont{Alford, Rajagopal,
  and Wilczek}}]{Alford:1998mk}
\bibinfo{author}{\bibfnamefont{M.~G.} \bibnamefont{Alford}},
  \bibinfo{author}{\bibfnamefont{K.}~\bibnamefont{Rajagopal}},
  \bibnamefont{and} \bibinfo{author}{\bibfnamefont{F.}~\bibnamefont{Wilczek}},
  \bibinfo{journal}{Nucl. Phys.} \textbf{\bibinfo{volume}{B537}},
  \bibinfo{pages}{443} (\bibinfo{year}{1999}), \eprint{hep-ph/9804403}.

\bibitem[{\citenamefont{Schmitt et~al.}(2004)\citenamefont{Schmitt, Wang, and
  Rischke}}]{Schmitt:2003aa}
\bibinfo{author}{\bibfnamefont{A.}~\bibnamefont{Schmitt}},
  \bibinfo{author}{\bibfnamefont{Q.}~\bibnamefont{Wang}}, \bibnamefont{and}
  \bibinfo{author}{\bibfnamefont{D.~H.} \bibnamefont{Rischke}},
  \bibinfo{journal}{Phys. Rev.} \textbf{\bibinfo{volume}{D69}},
  \bibinfo{pages}{094017} (\bibinfo{year}{2004}), \eprint{nucl-th/0311006}.

\bibitem[{\citenamefont{Iida and Baym}(2002{\natexlab{a}})}]{Iida:2002ev}
\bibinfo{author}{\bibfnamefont{K.}~\bibnamefont{Iida}} \bibnamefont{and}
  \bibinfo{author}{\bibfnamefont{G.}~\bibnamefont{Baym}},
  \bibinfo{journal}{Phys. Rev.} \textbf{\bibinfo{volume}{D66}},
  \bibinfo{pages}{014015} (\bibinfo{year}{2002}{\natexlab{a}}),
  \eprint{hep-ph/0204124}.

\bibitem[{\citenamefont{Iida}(2005)}]{Iida:2004if}
\bibinfo{author}{\bibfnamefont{K.}~\bibnamefont{Iida}}, \bibinfo{journal}{Phys.
  Rev.} \textbf{\bibinfo{volume}{D71}}, \bibinfo{pages}{054011}
  (\bibinfo{year}{2005}), \eprint{hep-ph/0412426}.

\bibitem[{\citenamefont{Alford and Sedrakian}(2010)}]{Alford:2010qf}
\bibinfo{author}{\bibfnamefont{M.~G.} \bibnamefont{Alford}} \bibnamefont{and}
  \bibinfo{author}{\bibfnamefont{A.}~\bibnamefont{Sedrakian}},
  \bibinfo{journal}{J. Phys.} \textbf{\bibinfo{volume}{G37}},
  \bibinfo{pages}{075202} (\bibinfo{year}{2010}), \eprint{1001.3346}.

\bibitem[{\citenamefont{Haber and Schmitt}(2018{\natexlab{a}})}]{Haber:2017oqb}
\bibinfo{author}{\bibfnamefont{A.}~\bibnamefont{Haber}} \bibnamefont{and}
  \bibinfo{author}{\bibfnamefont{A.}~\bibnamefont{Schmitt}},
  \bibinfo{journal}{J. Phys. G} \textbf{\bibinfo{volume}{45}},
  \bibinfo{pages}{065001} (\bibinfo{year}{2018}{\natexlab{a}}),
  \eprint{1712.08587}.

\bibitem[{\citenamefont{Son and Stephanov}(2008)}]{Son:2007ny}
\bibinfo{author}{\bibfnamefont{D.~T.} \bibnamefont{Son}} \bibnamefont{and}
  \bibinfo{author}{\bibfnamefont{M.~A.} \bibnamefont{Stephanov}},
  \bibinfo{journal}{Phys. Rev.} \textbf{\bibinfo{volume}{D77}},
  \bibinfo{pages}{014021} (\bibinfo{year}{2008}), \eprint{0710.1084}.

\bibitem[{\citenamefont{Alford and Good}(2008)}]{Alford:2007np}
\bibinfo{author}{\bibfnamefont{M.~G.} \bibnamefont{Alford}} \bibnamefont{and}
  \bibinfo{author}{\bibfnamefont{G.}~\bibnamefont{Good}},
  \bibinfo{journal}{Phys. Rev.} \textbf{\bibinfo{volume}{B78}},
  \bibinfo{pages}{024510} (\bibinfo{year}{2008}), \eprint{0712.1810}.

\bibitem[{\citenamefont{Haber and Schmitt}(2017{\natexlab{a}})}]{Haber:2016ljn}
\bibinfo{author}{\bibfnamefont{A.}~\bibnamefont{Haber}} \bibnamefont{and}
  \bibinfo{author}{\bibfnamefont{A.}~\bibnamefont{Schmitt}},
  \bibinfo{journal}{EPJ Web Conf.} \textbf{\bibinfo{volume}{137}},
  \bibinfo{pages}{09003} (\bibinfo{year}{2017}{\natexlab{a}}),
  \eprint{1612.01865}.

\bibitem[{\citenamefont{Haber and Schmitt}(2017{\natexlab{b}})}]{Haber:2017kth}
\bibinfo{author}{\bibfnamefont{A.}~\bibnamefont{Haber}} \bibnamefont{and}
  \bibinfo{author}{\bibfnamefont{A.}~\bibnamefont{Schmitt}},
  \bibinfo{journal}{Phys. Rev.} \textbf{\bibinfo{volume}{D95}},
  \bibinfo{pages}{116016} (\bibinfo{year}{2017}{\natexlab{b}}),
  \eprint{1704.01575}.

\bibitem[{\citenamefont{Haber and Schmitt}(2018{\natexlab{b}})}]{Haber:2018tqw}
\bibinfo{author}{\bibfnamefont{A.}~\bibnamefont{Haber}} \bibnamefont{and}
  \bibinfo{author}{\bibfnamefont{A.}~\bibnamefont{Schmitt}},
  \bibinfo{journal}{PoS} \textbf{\bibinfo{volume}{Confinement2018}},
  \bibinfo{pages}{213} (\bibinfo{year}{2018}{\natexlab{b}}),
  \eprint{1811.12302}.

\bibitem[{\citenamefont{Buckley et~al.}(2004)\citenamefont{Buckley, Metlitski,
  and Zhitnitsky}}]{Buckley:2004ca}
\bibinfo{author}{\bibfnamefont{K.~B.~W.} \bibnamefont{Buckley}},
  \bibinfo{author}{\bibfnamefont{M.~A.} \bibnamefont{Metlitski}},
  \bibnamefont{and} \bibinfo{author}{\bibfnamefont{A.~R.}
  \bibnamefont{Zhitnitsky}}, \bibinfo{journal}{Phys. Rev.}
  \textbf{\bibinfo{volume}{C69}}, \bibinfo{pages}{055803}
  (\bibinfo{year}{2004}), \eprint{hep-ph/0403230}.

\bibitem[{\citenamefont{Blaschke et~al.}(1999)\citenamefont{Blaschke,
  Sedrakian, and Shahabasian}}]{Blaschke:1999fy}
\bibinfo{author}{\bibfnamefont{D.}~\bibnamefont{Blaschke}},
  \bibinfo{author}{\bibfnamefont{D.}~\bibnamefont{Sedrakian}},
  \bibnamefont{and}
  \bibinfo{author}{\bibfnamefont{K.}~\bibnamefont{Shahabasian}},
  \bibinfo{journal}{Astron. Astrophys.} \textbf{\bibinfo{volume}{350}},
  \bibinfo{pages}{L47} (\bibinfo{year}{1999}), \eprint{astro-ph/9904395}.

\bibitem[{\citenamefont{Iida and Baym}(2001)}]{Iida:2000ha}
\bibinfo{author}{\bibfnamefont{K.}~\bibnamefont{Iida}} \bibnamefont{and}
  \bibinfo{author}{\bibfnamefont{G.}~\bibnamefont{Baym}},
  \bibinfo{journal}{Phys. Rev.} \textbf{\bibinfo{volume}{D63}},
  \bibinfo{pages}{074018} (\bibinfo{year}{2001}), \eprint{hep-ph/0011229}.

\bibitem[{\citenamefont{Iida and Baym}(2002{\natexlab{b}})}]{Iida:2001pg}
\bibinfo{author}{\bibfnamefont{K.}~\bibnamefont{Iida}} \bibnamefont{and}
  \bibinfo{author}{\bibfnamefont{G.}~\bibnamefont{Baym}},
  \bibinfo{journal}{Phys. Rev.} \textbf{\bibinfo{volume}{D65}},
  \bibinfo{pages}{014022} (\bibinfo{year}{2002}{\natexlab{b}}),
  \eprint{hep-ph/0108149}.

\bibitem[{\citenamefont{Sedrakian and Blaschke}(2002)}]{Sedrakian:2002mk}
\bibinfo{author}{\bibfnamefont{D.}~\bibnamefont{Sedrakian}} \bibnamefont{and}
  \bibinfo{author}{\bibfnamefont{D.}~\bibnamefont{Blaschke}},
  \bibinfo{journal}{Astrophysics} \textbf{\bibinfo{volume}{45}},
  \bibinfo{pages}{166} (\bibinfo{year}{2002}), \eprint{hep-ph/0205107}.

\bibitem[{\citenamefont{Giannakis and Ren}(2003)}]{Giannakis:2003am}
\bibinfo{author}{\bibfnamefont{I.}~\bibnamefont{Giannakis}} \bibnamefont{and}
  \bibinfo{author}{\bibfnamefont{H.-c.} \bibnamefont{Ren}},
  \bibinfo{journal}{Nucl. Phys.} \textbf{\bibinfo{volume}{B669}},
  \bibinfo{pages}{462} (\bibinfo{year}{2003}), \eprint{hep-ph/0305235}.

\bibitem[{\citenamefont{Hatsuda et~al.}(2006)\citenamefont{Hatsuda, Tachibana,
  Yamamoto, and Baym}}]{Hatsuda:2006ps}
\bibinfo{author}{\bibfnamefont{T.}~\bibnamefont{Hatsuda}},
  \bibinfo{author}{\bibfnamefont{M.}~\bibnamefont{Tachibana}},
  \bibinfo{author}{\bibfnamefont{N.}~\bibnamefont{Yamamoto}}, \bibnamefont{and}
  \bibinfo{author}{\bibfnamefont{G.}~\bibnamefont{Baym}},
  \bibinfo{journal}{Phys. Rev. Lett.} \textbf{\bibinfo{volume}{97}},
  \bibinfo{pages}{122001} (\bibinfo{year}{2006}), \eprint{hep-ph/0605018}.

\bibitem[{\citenamefont{Iida et~al.}(2004)\citenamefont{Iida, Matsuura,
  Tachibana, and Hatsuda}}]{Iida:2003cc}
\bibinfo{author}{\bibfnamefont{K.}~\bibnamefont{Iida}},
  \bibinfo{author}{\bibfnamefont{T.}~\bibnamefont{Matsuura}},
  \bibinfo{author}{\bibfnamefont{M.}~\bibnamefont{Tachibana}},
  \bibnamefont{and} \bibinfo{author}{\bibfnamefont{T.}~\bibnamefont{Hatsuda}},
  \bibinfo{journal}{Phys. Rev. Lett.} \textbf{\bibinfo{volume}{93}},
  \bibinfo{pages}{132001} (\bibinfo{year}{2004}), \eprint{hep-ph/0312363}.

\bibitem[{\citenamefont{Iida et~al.}(2005)\citenamefont{Iida, Matsuura,
  Tachibana, and Hatsuda}}]{Iida:2004cj}
\bibinfo{author}{\bibfnamefont{K.}~\bibnamefont{Iida}},
  \bibinfo{author}{\bibfnamefont{T.}~\bibnamefont{Matsuura}},
  \bibinfo{author}{\bibfnamefont{M.}~\bibnamefont{Tachibana}},
  \bibnamefont{and} \bibinfo{author}{\bibfnamefont{T.}~\bibnamefont{Hatsuda}},
  \bibinfo{journal}{Phys. Rev. D} \textbf{\bibinfo{volume}{71}},
  \bibinfo{pages}{054003} (\bibinfo{year}{2005}), \eprint{hep-ph/0411356}.

\bibitem[{\citenamefont{Schmitt et~al.}(2011)\citenamefont{Schmitt, Stetina,
  and Tachibana}}]{Schmitt:2010pf}
\bibinfo{author}{\bibfnamefont{A.}~\bibnamefont{Schmitt}},
  \bibinfo{author}{\bibfnamefont{S.}~\bibnamefont{Stetina}}, \bibnamefont{and}
  \bibinfo{author}{\bibfnamefont{M.}~\bibnamefont{Tachibana}},
  \bibinfo{journal}{Phys. Rev.} \textbf{\bibinfo{volume}{D83}},
  \bibinfo{pages}{045008} (\bibinfo{year}{2011}), \eprint{1010.4243}.

\bibitem[{\citenamefont{Ferrer et~al.}(2005)\citenamefont{Ferrer, de~la Incera,
  and Manuel}}]{Ferrer:2005vd}
\bibinfo{author}{\bibfnamefont{E.~J.} \bibnamefont{Ferrer}},
  \bibinfo{author}{\bibfnamefont{V.}~\bibnamefont{de~la Incera}},
  \bibnamefont{and} \bibinfo{author}{\bibfnamefont{C.}~\bibnamefont{Manuel}},
  \bibinfo{journal}{Phys. Rev. Lett.} \textbf{\bibinfo{volume}{95}},
  \bibinfo{pages}{152002} (\bibinfo{year}{2005}), \eprint{hep-ph/0503162}.

\bibitem[{\citenamefont{Ferrer et~al.}(2006)\citenamefont{Ferrer, de~la Incera,
  and Manuel}}]{Ferrer:2006vw}
\bibinfo{author}{\bibfnamefont{E.~J.} \bibnamefont{Ferrer}},
  \bibinfo{author}{\bibfnamefont{V.}~\bibnamefont{de~la Incera}},
  \bibnamefont{and} \bibinfo{author}{\bibfnamefont{C.}~\bibnamefont{Manuel}},
  \bibinfo{journal}{Nucl. Phys.} \textbf{\bibinfo{volume}{B747}},
  \bibinfo{pages}{88} (\bibinfo{year}{2006}), \eprint{hep-ph/0603233}.

\bibitem[{\citenamefont{Noronha and Shovkovy}(2007)}]{Noronha:2007wg}
\bibinfo{author}{\bibfnamefont{J.~L.} \bibnamefont{Noronha}} \bibnamefont{and}
  \bibinfo{author}{\bibfnamefont{I.~A.} \bibnamefont{Shovkovy}},
  \bibinfo{journal}{Phys. Rev.} \textbf{\bibinfo{volume}{D76}},
  \bibinfo{pages}{105030} (\bibinfo{year}{2007}), \bibinfo{note}{[Erratum:
  Phys. Rev.D86,049901(2012)]}, \eprint{0708.0307}.

\bibitem[{\citenamefont{Fukushima and Warringa}(2008)}]{Fukushima:2007fc}
\bibinfo{author}{\bibfnamefont{K.}~\bibnamefont{Fukushima}} \bibnamefont{and}
  \bibinfo{author}{\bibfnamefont{H.~J.} \bibnamefont{Warringa}},
  \bibinfo{journal}{Phys. Rev. Lett.} \textbf{\bibinfo{volume}{100}},
  \bibinfo{pages}{032007} (\bibinfo{year}{2008}), \eprint{0707.3785}.

\bibitem[{\citenamefont{Yu and Shovkovy}(2012)}]{Yu:2012jn}
\bibinfo{author}{\bibfnamefont{L.}~\bibnamefont{Yu}} \bibnamefont{and}
  \bibinfo{author}{\bibfnamefont{I.~A.} \bibnamefont{Shovkovy}},
  \bibinfo{journal}{Phys. Rev. D} \textbf{\bibinfo{volume}{85}},
  \bibinfo{pages}{085022} (\bibinfo{year}{2012}), \eprint{1202.0872}.

\bibitem[{\citenamefont{Eto et~al.}(2014)\citenamefont{Eto, Hirono, Nitta, and
  Yasui}}]{Eto:2013hoa}
\bibinfo{author}{\bibfnamefont{M.}~\bibnamefont{Eto}},
  \bibinfo{author}{\bibfnamefont{Y.}~\bibnamefont{Hirono}},
  \bibinfo{author}{\bibfnamefont{M.}~\bibnamefont{Nitta}}, \bibnamefont{and}
  \bibinfo{author}{\bibfnamefont{S.}~\bibnamefont{Yasui}},
  \bibinfo{journal}{PTEP} \textbf{\bibinfo{volume}{2014}},
  \bibinfo{pages}{012D01} (\bibinfo{year}{2014}), \eprint{1308.1535}.

\bibitem[{\citenamefont{Eto and Nitta}(2009)}]{Eto:2009kg}
\bibinfo{author}{\bibfnamefont{M.}~\bibnamefont{Eto}} \bibnamefont{and}
  \bibinfo{author}{\bibfnamefont{M.}~\bibnamefont{Nitta}},
  \bibinfo{journal}{Phys. Rev.} \textbf{\bibinfo{volume}{D80}},
  \bibinfo{pages}{125007} (\bibinfo{year}{2009}), \eprint{0907.1278}.

\bibitem[{\citenamefont{Alford et~al.}(2016)\citenamefont{Alford, Mallavarapu,
  Vachaspati, and Windisch}}]{Alford:2016dco}
\bibinfo{author}{\bibfnamefont{M.~G.} \bibnamefont{Alford}},
  \bibinfo{author}{\bibfnamefont{S.~K.} \bibnamefont{Mallavarapu}},
  \bibinfo{author}{\bibfnamefont{T.}~\bibnamefont{Vachaspati}},
  \bibnamefont{and} \bibinfo{author}{\bibfnamefont{A.}~\bibnamefont{Windisch}},
  \bibinfo{journal}{Phys. Rev.} \textbf{\bibinfo{volume}{C93}},
  \bibinfo{pages}{045801} (\bibinfo{year}{2016}), \eprint{1601.04656}.

\bibitem[{\citenamefont{Vachaspati and Achucarro}(1991)}]{Vachaspati:1991dz}
\bibinfo{author}{\bibfnamefont{T.}~\bibnamefont{Vachaspati}} \bibnamefont{and}
  \bibinfo{author}{\bibfnamefont{A.}~\bibnamefont{Achucarro}},
  \bibinfo{journal}{Phys. Rev.} \textbf{\bibinfo{volume}{D44}},
  \bibinfo{pages}{3067} (\bibinfo{year}{1991}).

\bibitem[{\citenamefont{Achucarro and Vachaspati}(2000)}]{Achucarro:1999it}
\bibinfo{author}{\bibfnamefont{A.}~\bibnamefont{Achucarro}} \bibnamefont{and}
  \bibinfo{author}{\bibfnamefont{T.}~\bibnamefont{Vachaspati}},
  \bibinfo{journal}{Phys. Rept.} \textbf{\bibinfo{volume}{327}},
  \bibinfo{pages}{347} (\bibinfo{year}{2000}), \eprint{hep-ph/9904229}.

\bibitem[{\citenamefont{Volkov}(2007)}]{Volkov:2006ug}
\bibinfo{author}{\bibfnamefont{M.~S.} \bibnamefont{Volkov}},
  \bibinfo{journal}{Phys. Lett. B} \textbf{\bibinfo{volume}{644}},
  \bibinfo{pages}{203} (\bibinfo{year}{2007}), \eprint{hep-th/0609112}.

\bibitem[{\citenamefont{Chernodub and Nedelin}(2010)}]{Chernodub:2010sg}
\bibinfo{author}{\bibfnamefont{M.~N.} \bibnamefont{Chernodub}}
  \bibnamefont{and} \bibinfo{author}{\bibfnamefont{A.~S.}
  \bibnamefont{Nedelin}}, \bibinfo{journal}{Phys. Rev.}
  \textbf{\bibinfo{volume}{D81}}, \bibinfo{pages}{125022}
  (\bibinfo{year}{2010}), \eprint{1005.3167}.

\bibitem[{\citenamefont{Bogomolny}(1976)}]{bogomolny}
\bibinfo{author}{\bibfnamefont{E.~B.} \bibnamefont{Bogomolny}},
  \bibinfo{journal}{Sov. J. Nucl. Phys. B} \textbf{\bibinfo{volume}{24}},
  \bibinfo{pages}{449} (\bibinfo{year}{1976}).

\bibitem[{\citenamefont{Bazeia et~al.}(2018{\natexlab{a}})\citenamefont{Bazeia,
  Marques, and Melnikov}}]{Bazeia:2018hyv}
\bibinfo{author}{\bibfnamefont{D.}~\bibnamefont{Bazeia}},
  \bibinfo{author}{\bibfnamefont{M.}~\bibnamefont{Marques}}, \bibnamefont{and}
  \bibinfo{author}{\bibfnamefont{D.}~\bibnamefont{Melnikov}},
  \bibinfo{journal}{Phys. Lett. B} \textbf{\bibinfo{volume}{785}},
  \bibinfo{pages}{454} (\bibinfo{year}{2018}{\natexlab{a}}),
  \eprint{1807.02007}.

\bibitem[{\citenamefont{Matthews et~al.}(1999)\citenamefont{Matthews, Anderson,
  Haljan, Hall, Wieman, and Cornell}}]{PhysRevLett.83.2498}
\bibinfo{author}{\bibfnamefont{M.~R.} \bibnamefont{Matthews}},
  \bibinfo{author}{\bibfnamefont{B.~P.} \bibnamefont{Anderson}},
  \bibinfo{author}{\bibfnamefont{P.~C.} \bibnamefont{Haljan}},
  \bibinfo{author}{\bibfnamefont{D.~S.} \bibnamefont{Hall}},
  \bibinfo{author}{\bibfnamefont{C.~E.} \bibnamefont{Wieman}},
  \bibnamefont{and} \bibinfo{author}{\bibfnamefont{E.~A.}
  \bibnamefont{Cornell}}, \bibinfo{journal}{Phys. Rev. Lett.}
  \textbf{\bibinfo{volume}{83}}, \bibinfo{pages}{2498} (\bibinfo{year}{1999}).

\bibitem[{\citenamefont{Rajagopal and Schmitt}(2006)}]{Rajagopal:2005dg}
\bibinfo{author}{\bibfnamefont{K.}~\bibnamefont{Rajagopal}} \bibnamefont{and}
  \bibinfo{author}{\bibfnamefont{A.}~\bibnamefont{Schmitt}},
  \bibinfo{journal}{Phys. Rev.} \textbf{\bibinfo{volume}{D73}},
  \bibinfo{pages}{045003} (\bibinfo{year}{2006}), \eprint{hep-ph/0512043}.

\bibitem[{\citenamefont{Tinkham}(2004)}]{tinkham2004introduction}
\bibinfo{author}{\bibfnamefont{M.}~\bibnamefont{Tinkham}},
  \emph{\bibinfo{title}{Introduction to Superconductivity}}
  (\bibinfo{publisher}{Dover Publications, New York}, \bibinfo{year}{2004}),
  ISBN \bibinfo{isbn}{9780486435039}.

\bibitem[{\citenamefont{Schmitt et~al.}(2002)\citenamefont{Schmitt, Wang, and
  Rischke}}]{Schmitt:2002sc}
\bibinfo{author}{\bibfnamefont{A.}~\bibnamefont{Schmitt}},
  \bibinfo{author}{\bibfnamefont{Q.}~\bibnamefont{Wang}}, \bibnamefont{and}
  \bibinfo{author}{\bibfnamefont{D.~H.} \bibnamefont{Rischke}},
  \bibinfo{journal}{Phys. Rev.} \textbf{\bibinfo{volume}{D66}},
  \bibinfo{pages}{114010} (\bibinfo{year}{2002}), \eprint{nucl-th/0209050}.

\bibitem[{\citenamefont{Ruester et~al.}(2005)\citenamefont{Ruester, Werth,
  Buballa, Shovkovy, and Rischke}}]{Ruester:2005jc}
\bibinfo{author}{\bibfnamefont{S.~B.} \bibnamefont{Ruester}},
  \bibinfo{author}{\bibfnamefont{V.}~\bibnamefont{Werth}},
  \bibinfo{author}{\bibfnamefont{M.}~\bibnamefont{Buballa}},
  \bibinfo{author}{\bibfnamefont{I.~A.} \bibnamefont{Shovkovy}},
  \bibnamefont{and} \bibinfo{author}{\bibfnamefont{D.~H.}
  \bibnamefont{Rischke}}, \bibinfo{journal}{Phys. Rev. D}
  \textbf{\bibinfo{volume}{72}}, \bibinfo{pages}{034004}
  (\bibinfo{year}{2005}), \eprint{hep-ph/0503184}.

\bibitem[{\citenamefont{Alford and Rajagopal}(2002)}]{Alford:2002kj}
\bibinfo{author}{\bibfnamefont{M.~G.} \bibnamefont{Alford}} \bibnamefont{and}
  \bibinfo{author}{\bibfnamefont{K.}~\bibnamefont{Rajagopal}},
  \bibinfo{journal}{JHEP} \textbf{\bibinfo{volume}{06}}, \bibinfo{pages}{031}
  (\bibinfo{year}{2002}), \eprint{hep-ph/0204001}.

\bibitem[{\citenamefont{Schmitt}(2010)}]{Schmitt:2010pn}
\bibinfo{author}{\bibfnamefont{A.}~\bibnamefont{Schmitt}},
  \bibinfo{journal}{Lect. Notes Phys.} \textbf{\bibinfo{volume}{811}},
  \bibinfo{pages}{1} (\bibinfo{year}{2010}), \eprint{1001.3294}.

\bibitem[{\citenamefont{Giannakis et~al.}(2004)\citenamefont{Giannakis, Hou,
  Ren, and Rischke}}]{Giannakis:2004xt}
\bibinfo{author}{\bibfnamefont{I.}~\bibnamefont{Giannakis}},
  \bibinfo{author}{\bibfnamefont{D.-f.} \bibnamefont{Hou}},
  \bibinfo{author}{\bibfnamefont{H.-c.} \bibnamefont{Ren}}, \bibnamefont{and}
  \bibinfo{author}{\bibfnamefont{D.~H.} \bibnamefont{Rischke}},
  \bibinfo{journal}{Phys. Rev. Lett.} \textbf{\bibinfo{volume}{93}},
  \bibinfo{pages}{232301} (\bibinfo{year}{2004}), \eprint{hep-ph/0406031}.

\bibitem[{\citenamefont{Haber}(2018)}]{Haber:2018yyd}
\bibinfo{author}{\bibfnamefont{A.}~\bibnamefont{Haber}}, Ph.D. thesis,
  \bibinfo{school}{TU Vienna} (\bibinfo{year}{2018}), \eprint{1811.12533}.

\bibitem[{\citenamefont{Fraga et~al.}(2019)\citenamefont{Fraga, Hippert, and
  Schmitt}}]{Fraga:2018cvr}
\bibinfo{author}{\bibfnamefont{E.~S.} \bibnamefont{Fraga}},
  \bibinfo{author}{\bibfnamefont{M.}~\bibnamefont{Hippert}}, \bibnamefont{and}
  \bibinfo{author}{\bibfnamefont{A.}~\bibnamefont{Schmitt}},
  \bibinfo{journal}{Phys. Rev. D} \textbf{\bibinfo{volume}{99}},
  \bibinfo{pages}{014046} (\bibinfo{year}{2019}), \eprint{1810.13226}.

\bibitem[{\citenamefont{Schmitt}(2020)}]{Schmitt:2020tac}
\bibinfo{author}{\bibfnamefont{A.}~\bibnamefont{Schmitt}},
  \bibinfo{journal}{Phys. Rev. D} \textbf{\bibinfo{volume}{101}},
  \bibinfo{pages}{074007} (\bibinfo{year}{2020}), \eprint{2002.01451}.

\bibitem[{\citenamefont{Bazeia et~al.}(2018{\natexlab{b}})\citenamefont{Bazeia,
  Marques, and Menezes}}]{Bazeia:2018eta}
\bibinfo{author}{\bibfnamefont{D.}~\bibnamefont{Bazeia}},
  \bibinfo{author}{\bibfnamefont{M.}~\bibnamefont{Marques}}, \bibnamefont{and}
  \bibinfo{author}{\bibfnamefont{R.}~\bibnamefont{Menezes}},
  \bibinfo{journal}{Phys. Rev. D} \textbf{\bibinfo{volume}{97}},
  \bibinfo{pages}{105024} (\bibinfo{year}{2018}{\natexlab{b}}),
  \eprint{1805.03250}.

\bibitem[{\citenamefont{Bazeia et~al.}(2018{\natexlab{c}})\citenamefont{Bazeia,
  Marques, and Olmo}}]{Bazeia:2018fhg}
\bibinfo{author}{\bibfnamefont{D.}~\bibnamefont{Bazeia}},
  \bibinfo{author}{\bibfnamefont{M.}~\bibnamefont{Marques}}, \bibnamefont{and}
  \bibinfo{author}{\bibfnamefont{G.~J.} \bibnamefont{Olmo}},
  \bibinfo{journal}{Phys. Rev. D} \textbf{\bibinfo{volume}{98}},
  \bibinfo{pages}{025017} (\bibinfo{year}{2018}{\natexlab{c}}),
  \eprint{1807.01299}.

\bibitem[{\citenamefont{Glampedakis et~al.}(2012)\citenamefont{Glampedakis,
  Jones, and Samuelsson}}]{Glampedakis:2012qp}
\bibinfo{author}{\bibfnamefont{K.}~\bibnamefont{Glampedakis}},
  \bibinfo{author}{\bibfnamefont{D.~I.} \bibnamefont{Jones}}, \bibnamefont{and}
  \bibinfo{author}{\bibfnamefont{L.}~\bibnamefont{Samuelsson}},
  \bibinfo{journal}{Phys. Rev. Lett.} \textbf{\bibinfo{volume}{109}},
  \bibinfo{pages}{081103} (\bibinfo{year}{2012}), \eprint{1204.3781}.

\bibitem[{\citenamefont{Penin and Weller}(2020)}]{Penin:2020cxj}
\bibinfo{author}{\bibfnamefont{A.~A.} \bibnamefont{Penin}} \bibnamefont{and}
  \bibinfo{author}{\bibfnamefont{Q.}~\bibnamefont{Weller}}
  (\bibinfo{year}{2020}), \eprint{2009.06640}.

\end{thebibliography}

\end{document}